\documentclass[10pt,a4paper]{article}
\usepackage[utf8]{inputenc}
\usepackage[english]{babel}
\usepackage{amsmath}
\usepackage{amsthm}
\numberwithin{equation}{section}
\usepackage{amsfonts}
\usepackage{mathtools}
\usepackage{mathrsfs}
\usepackage{mathpazo}
\usepackage{amssymb}
\usepackage{graphicx}
\usepackage{ifpdf}
\usepackage{braket}
\newcommand{\changefont}[3]{
\fontfamily{#1} \fontseries{#2} \fontshape{#3} \selectfont}
\changefont{ppl}{m}{n}

\usepackage{a4wide}

\usepackage[bookmarks=true,colorlinks=true,linkcolor=black,citecolor=black,urlcolor=black,bookmarksnumbered]{hyperref}

\theoremstyle{plain}

\theoremstyle{definition}

\theoremstyle{remark}

\def\ads{{\rm AdS}_5\times {\rm S}^5}

\expandafter\def\expandafter\bfseries\expandafter{\bfseries\ifmmode\else\boldmath\fi}
\expandafter\def\expandafter\mdseries\expandafter{\mdseries\ifmmode\else\unboldmath\fi}
\expandafter\def\expandafter\normalfont\expandafter{\normalfont\ifmmode\else\unboldmath\fi}

\begin{document}

\setcounter{equation}{0}
\setcounter{footnote}{0}
\setcounter{section}{0}

\thispagestyle{empty}

\begin{flushright} \texttt{HU-EP-16/29\\ HU-MATH-16/--}\end{flushright}

\begin{center}
\vspace{1.5truecm}

{\LARGE \bf Abelian Yang-Baxter Deformations \\and $TsT$ transformations}

\vspace{1.5truecm}

{David Osten and Stijn J. van Tongeren}

\vspace{1.0truecm}

{\em Institut f\"ur Mathematik und Institut f\"ur Physik, Humboldt-Universit\"at zu Berlin, \\ IRIS Geb\"aude, Zum Grossen Windkanal 6, 12489 Berlin, Germany}

\vspace{1.0truecm}

{{\tt osten@physik.hu-berlin.de, \quad svantongeren@physik.hu-berlin.de}}

\vspace{1.0truecm}
\end{center}

\begin{abstract} We prove that abelian Yang-Baxter deformations of superstring coset $\sigma$ models are equivalent to sequences of commuting $TsT$ transformations, meaning $T$ dualities and coordinate shifts. Our results extend also to fermionic deformations and fermionic $T$ duality, and naturally lead to a $TsT$ subgroup of the superduality group OSp$(d_b,d_b|2d_f)$. In cases like $\ads$, fermionic deformations necessarily lead to complex models. As an illustration of inequivalent deformations, we give all six abelian deformations of $\mathrm{AdS}_3$. We comment on the possible dual field theory interpretation of these (super-)$TsT$ models.
\end{abstract}

\newpage

\setcounter{equation}{0}
\setcounter{footnote}{0}
\setcounter{section}{0}

\tableofcontents

\section{Introduction}

Integrability is a key feature of the string $\sigma$ model on $\ads$ in the context of the AdS/CFT correspondence \cite{Maldacena.LargeNLimit}. Progress in this field has led to substantial improvements in our understanding of both sides of this duality \cite{Arutyunov.Frolov.Review,Beisert.Review,Bombardelli.et.al.Integrability.GaugeGravity}. One way to further extend our understanding is to study deformations that extend beyond the maximally symmetric example of $\ads$ and its lower dimensional cousins, while preserving integrability. The primary example of this is a string on the Lunin-Maldacena background \cite{Lunin.Maldacena.Deformation,Frolov:2005ty,Frolov.LaxPair.Lunin-Maldacena.Background}, dual to real $\beta$ deformed planar SYM. On the string side, this theory can be obtained by so-called $TsT$ transformations -- sequences of $T$ dualities and shifts in commuting directions, also known as Melvin twists. More recently it was realised in the manifestly integrability preserving framework of Yang-Baxter deformations. The purpose of this paper is to elucidate the connection between these two approaches.

Yang-Baxter $\sigma$ models were introduced as deformations of principal chiral models based on $R$ operators solving the modified classical Yang-Baxter equation \cite{Klimcik:2002zj}, preserving their integrability \cite{Klimcik.YBSM.Integrability}. This notion was generalised to symmetric space coset $\sigma$ models in \cite{DMV2} and then further to the supercoset $\sigma$ model describing the $\ads$ superstring \cite{DMV1}.\footnote{\label{footnote:lambdapoisson} These models are related to another type of integrable deformation known as the $\lambda$ model \cite{Sfetsos:2013wia,Hollowood:2014qma,Demulder:2015lva} by analytic continuation and Poisson-Lie duality \cite{Klimcik:1995ux,Klimcik:1995jn,Vicedo:2015pna,Hoare:2015gda,Sfetsos:2015nya,Klimcik:2015gba}, see also \cite{Delduc:2016ihq}. The $\lambda$-type models do correspond to solutions of supergravity \cite{Borsato:2016zcf,Chervonyi:2016ajp}.} By a simple limit this deformation procedure can be extended to solutions of the classical Yang-Baxter equation \cite{Kawaguchi:2014qwa}. These equations admit many solutions, and correspondingly there are many different integrable deformations of the $\ads$ string. In terms of general structure, at the level of symmetries, deformations based on the modified classical Yang-Baxter equation correspond to quantum deformations \cite{Delduc:2014kha}, while deformations based on the classical Yang-Baxter equation result in Drinfeld twists \cite{vanTongeren:2015uha}, see also \cite{Vicedo:2015pna}. At the level of string theory, the condition that the backgrounds of these models solve the supergravity equations of motion requires the associated $R$ operator to be unimodular \cite{Borsato.Wulff2016}. All Yang-Baxter deformations of the string preserve $\kappa$ symmetry however \cite{DMV1,Borsato.Wulff2016}, meaning that their backgrounds necessarily solve a set of modified supergravity equations \cite{Arutyunov:2015mqj,Wulff:2016tju}, guaranteeing scale but not Weyl invariance.

The structure described above matches with previously established results. Namely, the $\eta$ deformation of the string -- based on the modified classical Yang-Baxter equation -- was originally constructed using a non-unimodular $R$ operator, and indeed the associated background does not solve the supergravity equations \cite{Arutyunov.Borsato.Frolov.eta-deformations}, but rather the modified ones \cite{Arutyunov:2015mqj}, see also \cite{Hoare:2015wia}. Still, alternative $R$ operators exist \cite{Delduc:2014kha,Hoare.Tongeren.NonSplit.Split}. These appear to give inequivalent backgrounds, yet the same S matrix \cite{Hoare.Tongeren.NonSplit.Split}. None of the studied $R$ operators is unimodular, however, and it is not known whether a unimodular one exists.\footnote{Here it is interesting to recall that the bosonic part of the maximally deformed $\eta$ model can be completed to a solution of supergravity, giving the so-called mirror model \cite{Arutynov:2014ota,Arutyunov:2014cra,Arutyunov:2014jfa}. Algebraically this maximal deformation limit corresponds to a contraction \cite{Pachol:2015mfa}. The mirror model is an integrable model itself, and is closely related to the direct contraction of the full $\eta$ model \cite{Arutyunov.Borsato.Frolov.eta-deformations}. In particular the S matrices of these models appear to match.}

For classical Yang-Baxter deformations the situation is more diverse. $R$ operators of this type can be divided into abelian and non-abelian, depending on whether the associated generators all mutually commute or not. In the non-abelian class, bosonic jordanian $R$ operators are not unimodular, and indeed the associated backgrounds solve the modified supergravity equations \cite{Hoare.Tongeren.Jordanian}, but not the regular ones \cite{Kyono.Yoshida.Yang-Baxter-deformation,Hoare.Tongeren.Jordanian}. In fact, many jordanian deformations are closely related to the $\eta$ model, as they can be obtained from it by singular boosts \cite{Hoare.Tongeren.Jordanian}. Further bosonic jordanian examples were recently investigated in \cite{Orlando:2016qqu}. The conformal symmetry of AdS$_5$ is large enough, however, to admit other, unimodular non-abelian $R$ operators \cite{Borsato.Wulff2016}.

In contrast to non-abelian ones, abelian $R$ operators are always unimodular, meaning any such operator maps a solution of supergravity to a solution of supergravity. Various abelian deformations were studied at the bosonic level, see e.g. \cite{Matsumoto:2014gwa,MatsumotoYoshida.Gravity-CYBE,Tongeren.CYBE-deformations,Matsumoto:2016lnr}, including the Lunin-Maldacena background mentioned above \cite{Matsumoto:2014nra}. More recently some examples have been studied to quadratic order in fermions, both as singular boosts of the $\eta$ model \cite{Arutyunov.Borsato.Frolov.eta-deformations,Hoare.Tongeren.Jordanian} and directly \cite{Kyono.Yoshida.Yang-Baxter-deformation}. These individual examples all fit the proposal of one of the present authors \cite{Tongeren.CYBE-deformations}, that abelian Yang-Baxter deformations are equivalent to sequences of commuting $TsT$ transformations.

The objective of this paper is to get closer to a complete understanding of this abelian class of Yang-Baxter deformations, by giving a general proof of the equivalence between abelian Yang-Baxter deformations and (sequences of commuting) $TsT$ transformations. This proof relies on always being able to find a group parameterisation such that the Maurer-Cartan forms manifest a set of chosen commuting isometries. For completeness, upon complexification we can extend our proof to include $R$ operators based on anticommuting supercharges. These are equivalent to a generalised fermionic version of $TsT$ transformations, which we introduce. Furthermore, in order to explore the various possible abelian deformations/$TsT$ transformations and to get a better idea of their general structure, we consider AdS$_3$ -- the simplest nontrivial non-compact example -- which admits six inequivalent abelian deformations.

This paper is organised as follows. In section \ref{chap:eta-deformation} we establish our conventions for the type IIB superstring in $\ads$ and its integrable deformations based on the classical Yang-Baxter equation. Bosonic and fermionic $T$ duality is introduced in section \ref{chap:TDuality}, where we also briefly discuss the duality groups O$(d,d)$ and OSp$(d_b,d_b|2d_f)$ respectively. We prove equivalence between abelian deformations and $TsT$ transformations in section \ref{chap:Supergravity}. In the last section we address the fact that there are different inequivalent commuting subalgebras in non-compact cosets, illustrating this with a discussion of all inequivalent abelian deformations of AdS$_3$. In the conclusions we indicate some open questions and comment on the possible dual field theory interpretation of these deformed models.


\section{Yang-Baxter Deformations}
\label{chap:eta-deformation}

\subsubsection*{The Undeformed $\ads$ Superstring Action}
Let us briefly introduce the conventions for the supercoset $\sigma$ model with fields in
\begin{equation}
\mathcal{M} = \frac{\text{PSU}(2,2|4)}{\text{SO}(1,4) \times \text{SO}(5)} \simeq AdS_5 \times S^5 \times \mathbb{C}^{0|16},
\end{equation}
which describes the Green-Schwarz type IIB superstring in $\ads$ \cite{Metsaev.Tseytlin.TypeIIbGSAction.AdS5S5}, see \cite{Arutyunov.Frolov.Review} for an extensive review. The argumentation in the section \ref{chap:Supergravity} will also hold for general bosonic symmetric space $\sigma$ models and any supercoset $\sigma$ models which can be described similarly to the $\ads$ superstring.

The string moving in a coset $\mathcal{M} = G/H$ is described by $G$ valued fields $g: \Sigma \rightarrow G$ defined on the worldsheet $\Sigma$. The theory can be formulated in terms of the Maurer-Cartan forms taking values in the Lie algebra $\mathfrak{g}$ of $G$
\begin{equation}
A = - g^{-1} \mathrm{d} g \in \mathfrak{g}.
\end{equation}
Important for the integrability of the $\ads$ superstring is the existence of the $\mathbb{Z}_4$-grading of $\mathfrak{g} = \mathfrak{su}(2,2|4)$:
\begin{equation}
\mathfrak{g} = \mathfrak{g}^{(0)} \oplus \mathfrak{g}^{(1)} \oplus \mathfrak{g}^{(2)} \oplus \mathfrak{g}^{(3)},
\end{equation}
with the properties
\begin{align*}
[M^{(i)} , N^{(j)} ] \in \mathfrak{g}^{(i+j\ \text{mod} \ 4)} \quad \text{for} \ M^{(k)},N^{(k)} \in \mathfrak{g}^{(k)},
\end{align*}
and for the supertrace of a matrix realisation of $\mathfrak{g}$
\begin{align*}
\text{STr}(M^{(i)} N^{(j)} ) = 0 \qquad \text{for} \ m + n \neq 0 \ \text{mod} \ 4.
\end{align*}
$\mathfrak{g}^{(2)}$ denotes the bosonic coset algebra, $\mathfrak{g}^{(0)}$ the little group algebra of the bosonic coset, and  $\mathfrak{g}^{(1)}$ and $\mathfrak{g}^{(3)}$ are the odd parts of the algebra.\footnote{We choose our superalgebra conventions as in \cite{Arutyunov.Frolov.Review}, where elements of the algebra may be represented as an \textit{even} supermatrix
\begin{equation}
\left( \begin{array}{cc} m & \eta \\ \vartheta & n \end{array} \right) \quad \text{with} \ m, n: \ \text{matrices built from $c$-numbers}, \ \eta, \vartheta \ \text{Grassmann-valued matrices}
\end{equation}
Let us note, that we work with bosonic generators $\{h_i\}$ and fermionic generators $\{Q_\alpha \}$ being even respectively odd supermatrices with only even entries, so that e. g.
\begin{align*}
g=\exp(X^i h_i + \theta^\alpha Q_\alpha) \qquad A = -g^{-1} \mathrm{d} g
\end{align*}
are even supermatrices for a Grassmann-valued fields $\theta^\alpha$.}

The action of the superstring in $\ads$ in conformal gauge\footnote{This is purely a choice of convenience and does not affect our analysis.} takes the form \cite{Metsaev.Tseytlin.TypeIIbGSAction.AdS5S5}
\begin{equation}
S \propto \int \mathrm{d}^2 \sigma \ \mathcal{L} = \int \mathrm{d}^2 \sigma \  \text{STr}(A_+ d_- (A_-) )  \label{eq:UndeformedLag},
\end{equation}
with the worldsheet light-cone components of $A$
\begin{align*}
A_\pm = A_M \partial_\pm Z^M,
\end{align*}
and the linear combinations of projection operators on the $\mathbb{Z}_4$-components
\begin{equation}
d_\pm = \mp \mathfrak{P}^{(1)} + 2 \mathfrak{P}^{(2)} \pm \mathfrak{P}^{(3)}.
\end{equation}
Key features of the $\sigma$ model \eqref{eq:UndeformedLag} are $\kappa$ symmetry and integrability. The latter is associated to a spectral parameter dependent Lax pair
\begin{equation}
{L}_\pm(\lambda) = A_\pm^{(0)} + \lambda A_\pm^{(1)} + \lambda^{\mp 2} A_\pm^{(2)} + \lambda^{-1} A_\pm^{(3)}, \label{eq:UndeformedLax}
\end{equation}
where the flatness condition
\begin{equation}
\partial_+ L_- - \partial_- L_+ - [ L_+ , L_- ] = 0
\end{equation}
is equivalent to the equations of motion.

Let us now introduce integrable deformations of (super)coset $\sigma$ models such as \eqref{eq:UndeformedLag}, based on solutions of the classical Yang-Baxter equation.

\subsubsection*{The Classical Yang-Baxter Equation and Linear $R$ operators}
The standard form of the classical Yang-Baxter equation (CYBE) defined on tensor products of an algebra or superalgebra $\mathfrak{g}$ is
\begin{align*}
[ r_{12} , r_{13} ] + [r_{12} , r_{23} ] + [r_{13} , r_{23} ] = 0 \qquad \text{for} \ r \in \mathfrak{g} \otimes \mathfrak{g} .
\end{align*}
Deformations are formulated in terms of equivalent linear operators $R: \mathfrak{g} \rightarrow \mathfrak{g}$. The transition from a graded skewsymmetric $r$ matrix to an $R$ operator is via the trace
\begin{align*}
r &= a \wedge b := \frac{1}{2}( a \otimes b - (-1)^{s(a)s(b)} b \otimes a) \\
\rightarrow \quad R(M) &:= \text{STr}_2(r \cdot (1 \otimes M) ) = \frac{1}{2} \left( a \text{STr}(bM) - (-1)^{s(a)s(b)} b \text{STr}(aM) \right),
\end{align*}
extended by linearity, where we refer to the parity of a supermatrix $a$ as $s(a)$. The CYBE in terms of the $R$ operator takes the form
\begin{equation}
[R(M) , R(N)] - R\left( [R(M),N] + [M,R(N)]\right) = 0 . \label{eq:CYBE}
\end{equation}
Note that for the parities of a $r$ matrix $r = a \wedge b$ and the associated $R$ operator we have $s(r) = s(R) = s(a)s(b)$ and $s(R(M)) = s(R)s(M)$.

A simple solution of \eqref{eq:CYBE} over a given algebra $\mathfrak{g}$ is the $r$ matrix consisting of graded commuting generators. In the following we will call these $r$ matrices \textit{abelian}.

\subsubsection*{Deformations based on Solutions of the Classical Yang-Baxter Equation}
Yang-Baxter deformations of coset $\sigma$ models of the form of eqn. \eqref{eq:UndeformedLag} are generated by skew-symmetric\footnote{This means STr$(M R(N))=-$STr$(R(M)N)$.} linear $R$ operators solving \eqref{eq:CYBE}. A further ingredient is the ``dressing'' of the $R$ operator $R_g = \text{Ad}_g^{-1} \circ R \circ \text{Ad}_g$. The Yang-Baxter deformed action is given by \cite{DMV1,Kawaguchi:2014qwa}
\begin{align}
S \propto \int \mathrm{d}^2 \sigma \ \mathcal{L} = \int \mathrm{d}^2 \sigma \ \text{STr} \left( A_+ d_- (J_-) \right), \label{eq:DeformedLagCYBE}
\end{align}
where we introduced the deformed currents $J_\pm = \frac{1}{\mathbb{1} \pm R_g \circ d_-}(A_\pm)$, and directly specified to the (unmodified) classical Yang-Baxter case. Note that we include deformation parameters already in the definition of $R$. These can take any real respectively Grassmannian value depending on the parity of the generating $R$ operator, as the CYBE \eqref{eq:CYBE} is invariant under rescalings of $R$.

These deformations preserve the $\kappa$ symmetry and integrability of the undeformed model \eqref{eq:UndeformedLag}. The associated deformed Lax pair is
\begin{equation}
L_\pm = J_\pm^{(0)} + \lambda J_\pm^{(1)} + \lambda^{\mp 2} J_\pm^{(2)} + \lambda^{-1} J_\pm^{(3)}. \label{eq:DeformedLaxCYBE}
\end{equation}
These deformations break part of the global $G$ symmetry $g \mapsto g^\prime g$ for $g^\prime \in G$ of the undeformed model. The unbroken symmetries are generated by the generators $T$ for which \cite{Tongeren.CYBE-deformations}
\begin{equation}
R([T, M]) = [T,R(M)]  \quad \forall M \in \mathfrak{g}\label{eq:DeformedSymmetryGenerators}.
\end{equation}

\section{$T$ Duality Groups and their $TsT$ Subgroups}
\label{chap:TDuality}

In this section we will briefly recall bosonic and fermionic $T$ duality and the associated $TsT$ transformations in the $\sigma$ model context.

\subsection{The Notion of Bosonic and Fermionic $T$ duality}
Consider a generic (classical\footnote{A dilaton $\phi$ enters the string action at a higher order in the coupling  $\alpha^\prime$. At the classical level the dilaton has to be introduced in the corresponding supergravity (e.g. the $RR$-forms appear always as $e^\phi F_{\mu_1 ... \mu_p}$). As we will not do explicit field redefinitions, we neglect it and its behaviour under $T$ duality from the start. Working at the classical level we also disregard any prefactors of the action and are only interested in its schematical form.}) string $\sigma$ model of the form
\begin{equation}
S  \propto \int \mathrm{d}^2 \sigma \ \partial_+ Z^M \mathcal{E}_{M N}(Z) \partial_- Z^N \equiv  \int \mathrm{d}^2 \sigma \mathcal{L}, \qquad M,N = 1, ... ,D \ , \label{eq:ActionSigma}
\end{equation}
where we work in conformal gauge for the sake of convenience, and understand $Z^M$ as
\begin{equation}
Z^M = ( X^\mu(\sigma) , \theta^{\Delta}(\sigma) ) \nonumber
\end{equation}
with some bosonic fields $X^\mu$ and some fermionic Grassmann-valued fields $\theta^{\Delta}$. We refer to the parity of the coordinate $Z^M$ as $s(M)$. $\mathcal{E}_{M N}(Z)$ is the background field describing the coupling between the fields\footnote{$\mathcal{E}_{M N}$ could be decomposed into its graded symmetric (metric-like) and graded skewsymmetric part: $\mathcal{E}_{M N} = \mathcal{G}_{MN} + \mathcal{B}_{MN}$. But only the order $\theta^0$ terms in $\mathcal{G}_{\mu\nu}$ respectively $\mathcal{B}_{\mu\nu}$ would have a direct physical interpretation as the components of metric and $B$ field. We stick to the quite abstract 'background' $\mathcal{E}_{MN}$ as it is practical and sufficient for our further considerations.} with parity $s(\mathcal{E}_{MN}) = s(M) + s(N)$, so that $s(\mathcal{L}) = 0$.

Now we assume the model has a manifest isometry and choose the associated coordinate to be $Z^1$, meaning the symmetry is realised as a shift of $Z^1$. We write $Z^M = (Z^1 , Z^{\underline{M}})$ with $\underline{M}=2,...,D$, so that $\mathcal{E}_{MN} \equiv \mathcal{E}_{MN}(Z^{\underline{M}})$. $Z^1$ can be either bosonic or fermionic\footnote{In the fermionic case the generator $Q$ dual to the isometry coordinate has to anticommute with itself in order to correspond to a shift isometry. In other words, fermionic $T$ duality requires a supercharge $Q$ with $Q^2=0$. We will come back to this point below.}. This allows us to rewrite the Lagrangian by introducing gauge fields $A_\pm$:
\begin{align*}
\partial_\pm Z^1 &\rightarrow A_\pm \qquad \mathcal{L} \rightarrow \mathcal{L} - \bar{Z}^1 (\partial_+ A_- - \partial_- A_+),
\end{align*}
where the Lagrange multiplier $\bar{Z}^1$ ensures $A_\pm = \partial_\pm Z^1$ by its equations of motion. Integrating out $A_\pm$ instead of $\bar{Z}^1$ yields the action of the dual model
\begin{equation}
\bar{S} \propto \int \mathrm{d}^2 \sigma \ \partial_+ \bar{X}^M  \bar{\mathcal{E}}_{M N}\partial_- \bar{X}^N , \nonumber
\end{equation}
with the dual background $\bar{\mathcal{E}}$ given by
\begin{align}
\bar{\mathcal{E}}_{11} &= (-1)^{s(1)} \frac{1}{\mathcal{E}_{11}}, \qquad \bar{\mathcal{E}}_{1 \underline{M}} = (-1)^{s(1)} \frac{\mathcal{E}_{1 \underline{M}}}{\mathcal{E}_{11}}, \qquad \bar{\mathcal{E}}_{\underline{M} 1} = - \frac{\mathcal{E}_{\underline{M}1}}{\mathcal{E}_{11}} \nonumber \\
\bar{\mathcal{E}}_{\underline{M} \ \underline{N}} &= \mathcal{E}_{\underline{M} \ \underline{N}}  - \frac{\mathcal{E}_{\underline{M}1} \mathcal{E}_{1 \underline{N}} }{\mathcal{E}_{11}} \qquad \qquad  \text{for} \ \underline{M},\underline{N}=2,...,D. \label{eq:BuscherRules}
\end{align}
For $T$ duality along a bosonic isometry we reproduce Buscher's $T$ duality rules \cite{Buscher}. For details on topological considerations and fermionic $T$ duality and its implications in general we refer to e.g. \cite{Berkovits.Maldacena.fermT-Duality,Beisert.Ricci.Tseytlin.Wolf.DualSuperconfSym}.\footnote{Note that our conventions for the $\sigma$ model \eqref{eq:ActionSigma} differ from \cite{Berkovits.Maldacena.fermT-Duality}, leading to some different signs in \eqref{eq:BuscherRules}. Furthermore note that, as defined, along a fermionic isometry coordinate only $T^4$, not $T^2$, is manifestly the identity operation. $T^2$ is a trivial and physically irrelevant coordinate redefinition of the background, $Z^1 \rightarrow (-1)^{s(1)}Z^1$, however.}

\subsection{The O$(d,d)$ Group of Bosonic $T$ duality}
Now we assume the model has $d$ commuting bosonic isometries and choose the associated coordinates to be $X^i$ for $i=1,...,d$. We write $Z^M = (X^i,Z^{\underline{i}})$ with the $Z^{\underline{i}}$ denoting the $D-d$ remaining non-isometry coordinates. In particular, $\mathcal{E}_{MN} \equiv \mathcal{E}_{MN}(Z^{\underline{i}})$. With the following fractional linear action of a $2D \times 2D$-matrix $G$ on $\mathcal{E}$
\begin{equation}
G = \left( \begin{array}{cc} A & B \\ C & D \end{array} \right) \quad \rightarrow \quad \tilde{\mathcal{E}} = (A \mathcal{E} + B ) (C\mathcal{E} + D)^{-1}, \label{eq:O(D,D)_fraclin}
\end{equation}
a $T$ duality transformation along $X^i$ can be represented for every $i \in \{1,...,d \}$ as
\begin{equation}
G_{T_i} = \left( \begin{array}{cc} \mathbb{1}_D - E_i & - E_i \\ - E_i & \mathbb{1}_D - E_i \end{array} \right), \label{eq:Ti_O(D,D)}
\end{equation}
where $E_i$ is the $D\times D$-matrix with every element being zero, except for $(E_i)_{ii} = 1$. Other transformations, that even leave the Lagrangian invariant, are GL$(d)$-transformations of the isometry directions if we also transform $\mathcal{E}$ accordingly. Let $A \in$GL$(d)$ and
\begin{align*}
X^i &\rightarrow \bar{X}^i = A^{ij} X^j, \qquad \qquad Z^{\underline{i}} \rightarrow Z^{\underline{i}},
\end{align*}
then the Lagrangian is invariant if
\begin{align*}
\tilde{\mathcal{E}} = \left( \begin{array}{cc} (A^T)^{-1} & \\ & \mathbb{1}_{D-d} \end{array} \right) \cdot \mathcal{E} \cdot \left( \begin{array}{cc} A^{-1} & \\ & \mathbb{1}_d \end{array} \right).
\end{align*}
This can be represented by fractional linear action \eqref{eq:O(D,D)_fraclin} on $\mathcal{E}$ of the group element
\begin{equation}
G_{GL} = \left( \begin{array}{cccc} (A^T)^{-1} & & & \\ & \mathbb{1}_{D-d} & & \\ & & A & \\ & & & \mathbb{1}_{D-d} \end{array} \right). \label{eq:GL(d)_O(D,D)}
\end{equation}
Both $G_{T_i}$ and $G_{GL}$ are elements of O$(D,D)$, where we understand its elements as $2D \times 2D$-matrices $G$ fulfilling the pseudo-orthogonality relation
\begin{equation}
G J G^T = J , \qquad \qquad J = \left( \begin{array}{cc} & \mathbb{1}_D \\ \mathbb{1}_D & \end{array} \right). \label{eq:PseudoorthBos}
\end{equation}
The form of \eqref{eq:Ti_O(D,D)} and \eqref{eq:GL(d)_O(D,D)} suggests that we can write these as elements of O$(d,d)$\footnote{From discussions of the spectrum one can motivate the $T$ duality group being the group of toroidal compactifications O$(d,d,\mathbb{Z})$. For example for closed strings, O$(d,d,\mathbb{Z})$ transformations correspond to ``rotations'' on the lattice describing winding numbers and Kaluza-Klein excitation numbers associated to the compact (toroidal) $($U$(1))^d$-isometry, which leave the spectrum invariant. This is reviewed in e.g. \cite{Giveon.et.al.Target.space.Duality}. In the above $\sigma$ model, however, we consider theories that are equivalent modulo boundary conditions; $TsT$ transformations can be absorbed in twisted boundary conditions \cite{Frolov.LaxPair.Lunin-Maldacena.Background,Alday:2005ww}.} embedded in O$(D,D)$
\begin{align}
g =  \left( \begin{array}{cc} a & b \\ c & d \end{array} \right) \in \text{O}(d,d) \quad \rightarrow \quad G =  \left( \begin{array}{cc|cc} a & & b & \\ &  \mathbb{1}_{D-d} & & 0_{D-d} \\  \hline c & & d & \\ & 0_{D-d}  & & \mathbb{1}_{D-d} \end{array} \right) \in \text{O}(D,D). \label{eq:O(d,d)embedding}
\end{align}
Note that $\det g_{T_i} = -1$, so in fact bosonic $T$ duality transformations itself are not in the component connected to the identity, in contrast to $g_{GL}$. But we can generate further elements of the component connected to the identity of O$(d,d)$ by a product of some general linear transformations and an even number of $T$ duality transformations.

\subsubsection*{Bosonic $TsT$ Transformations}
Now we introduce $TsT$ transformations in the above framework. These gained some attention in the context of the AdS/CFT correspondence, as a particular $TsT$ transformation of the $\ads$ background gives a supergravity background dual to $\beta$ deformed SYM \cite{Lunin.Maldacena.Deformation}. To do $TsT$ transformations we need at least two isometries, which we parameterise by $X^1$ and $X^2$ in the following. A single $TsT$ transformation is generated by a $T$ duality transformation on the $X^1$, a shift\footnote{Note that this a quite specific transformation. Generic coordinate transformations would also lead to contributions in the other blocks of an O$(d,d)$ element in comparison to \eqref{eq:singleTsT}). Shifts in the ``other'' direction like
\begin{equation}
\bar{X}^1 \rightarrow \bar{X}^1 - \theta \bar{X}^2 \label{eq:ThetaShift-coordinate}
\end{equation}
between two $T$ duality transformations would lead to
\begin{equation}
g_{\Theta_{12}} =  \left( \begin{array}{cccc} 1 & & 0 & -\theta \\ & 1 & \theta & 0 \\  & & 1 & \\ & & & 1 \end{array} \right) ,
\end{equation}
these are called $\Theta$ shifts and build an abelian subgroup of O$(d,d)$, created by skewsymmetric $d \times d$-matrices $\Theta$ in the upper right block:
\begin{equation}
g_\Theta = \left( \begin{array}{cc} \mathbb{1}_d & \Theta \\  & \mathbb{1}_d \end{array} \right) \in \text{SO}(d,d). \label{eq:ThetaShift_SO(d,d)}
\end{equation}
The background is transformed with \eqref{eq:O(d,d)embedding} and \eqref{eq:O(D,D)_fraclin} only in the isometry components as
\begin{equation}
\tilde{\mathcal{E}}_{ij} = \mathcal{E}_{ij} + \Theta_{ij}  \qquad \leftrightarrow \qquad \bar{B}_{ij} = B_{ij} + \Theta_{ij} \nonumber,
\end{equation}
where $B_{ij}$ are components corresponding to the isometry directions of the $B$-field. While these coordinate shifts \eqref{eq:ThetaShift-coordinate} look quite similar to the ones of $TsT$ transformations, $\Theta$ shifts act very differently on the background. $\Theta$ shifts clearly generate physically equivalent models up to boundary terms, as $H=dB$ remains invariant.}
\begin{align}
\bar{X}_2 \rightarrow \bar{X}_2 - \gamma \bar{X}_1 \label{eq:shift_TsT}
\end{align}
and then a $T$ duality transformation on the $\bar{X}^1$ direction back. In the above group language, in the minimal $d=2$ setting this looks like
\begin{equation}
g_{\Gamma_{12}} = g_{T_1} \cdot \left( \begin{array}{cccc} 1 & \gamma & & \\ 0 & 1 & & \\ & & 1 & 0 \\ & & -\gamma & 1 \end{array} \right) \cdot  g_{T_1} =  \left( \begin{array}{cccc} 1 &  & & \\ & 1 & & \\ 0 & - \gamma & 1 & \\ \gamma & 0 & & 1 \end{array} \right) .\label{eq:singleTsT}
\end{equation}
Generic $TsT$ transformations can be understood as the straightforward generalisation to fractional linear transformations of the type \eqref{eq:O(D,D)_fraclin} with the generating group element
\begin{equation}
 g_\Gamma = \left( \begin{array}{cc} \mathbb{1}_d & \\ \Gamma & \mathbb{1}_d \end{array} \right) \in \text{SO}(d,d), \label{eq:TsT_SO(d,d)}
\end{equation}
where $\Gamma$ is an antisymmetric $d\times d$-matrix. This can be seen as
\begin{equation}
g_{\Gamma_1} \cdot  g_{\Gamma_2} =  \left( \begin{array}{cc} \mathbb{1}_d & \\ \Gamma_1 + \Gamma_2 & \mathbb{1}_d \end{array} \right) =  g_{\Gamma_1 + \Gamma_2} \label{eq:TsTSO(d,d)closure},
\end{equation}
meaning we can construct generic $TsT$ transformations by executing subsequent single $TsT$ transformation. $TsT$ transformations form an abelian subgroup of the component connected to the identity of O$(d,d)$.

\subsection{OSp$(d_b,d_b|2 d_f)$ as the Superduality Group}
Consider a background $\mathcal{E}$ with $d_b$ bosonic and $d_f$ fermionic isometries and $d = d_b + d_f$. Let us write our coordinates as
\begin{equation}
Z^M = (Z^a , Z^{\underline{a}}) = (X^i , \theta^\alpha , Z^{\underline{a}}), \qquad \text{with} \ i=1,...,d_b \ \text{and} \ \alpha = 1,...,d_f. \label{eq:CoordFerm}
\end{equation}
The matrix representation in the sense of \eqref{eq:O(D,D)_fraclin} and \eqref{eq:O(d,d)embedding} of a single $T$ duality transformation \eqref{eq:BuscherRules} along the isometry coordinate $Z^a$ is\footnote{Note that $\det g_{T_a} = - (-1)^{s(a)}$.}
\begin{equation}
g_{T_a} = \left( \begin{array}{cc} \mathbb{1}_d - E_a & - E_a \\ - (-1)^{s(a)} E_a & \mathbb{1}_d - E_a \end{array} \right) \label{eq:Tferm}.
\end{equation}
We can further consider GL$(d_b|d_f)$ coordinate transformations of the $Z^a= (X^i , \theta^\alpha)$
\begin{align*}
Z^a \rightarrow \bar{Z}^a = {A^a}_b Z^b
\end{align*}
with a supermatrix
\begin{equation}
A = \left( \begin{array}{cc} m & \eta \\ \vartheta & n \end{array} \right) \in \text{GL}(d_b|d_f). \nonumber
\end{equation}
With supertransposition defined as
\begin{equation}
A^{ST} = \left( \begin{array}{cc} m & \eta \\ \vartheta & n \end{array} \right)^{ST} = \left( \begin{array}{cc} m^T & \vartheta^T \\ -\eta^T & n^T \end{array} \right) \nonumber,
\end{equation}
the ``group element'' of such a GL$(d_b|d_f)$-transformation with the action \eqref{eq:O(D,D)_fraclin} on the background components $\mathcal{E}$ in the conventions of \eqref{eq:ActionSigma} is given similarly to \eqref{eq:GL(d)_O(D,D)} by
\begin{equation}
g_{GL} = \left( \begin{array}{cc} (A^{ST})^{-1} & \\ & A \end{array} \right) \qquad
\text{for} \ A \in \text{GL}(d_b|d_f). \label{eq:GLferm}
\end{equation}
It is easy to show that both \eqref{eq:Tferm} and \eqref{eq:GLferm} are elements of a group with elements
\begin{align*}
g = \left( \begin{array}{cc} A & B \\ C & D \end{array} \right) \quad \text{with} \ A,B,C,D \in \mathbb{R}^{(d_b|d_f)\times (d_b|d_f)}
\end{align*}
fulfilling a modified pseudoorthogonality relation (in comparison to \eqref{eq:PseudoorthBos})
\begin{equation}
g J g^{ST} = J \qquad \text{with} \ \left( \begin{array}{cc} A & B \\ C & D \end{array} \right)^{ST} := \left( \begin{array}{cc} A^{ST} & C^{ST} \\ B^{ST} & D^{ST} \end{array} \right) \quad \text{and} \quad \small{J = \left( \begin{array}{cccc} & & \mathbb{1}_{d_b} & \\ & & & \mathbb{1}_{d_f} \\ \mathbb{1}_{d_b} & & & \\ & -\mathbb{1}_{d_f} & & \end{array} \right)}. \label{eq:TDualityGroupFerm}
\end{equation}
This is a representation\footnote{More commonly one defines OSp$(m,m|2n)$ as the group constisting of $(2m|2n)\times (2m|2n)$-supermatrices $M$ preserving the supermetric $\mathcal{J}$
\begin{equation}
M \mathcal{J} M^{ST} = \mathcal{J} \quad \text{with} \ \mathcal{J} = \left( \begin{array}{cc|cc} \mathbb{1}_m & & & \\ & -\mathbb{1}_m & & \\ \hline & & & \mathbb{1}_n \\ & & -\mathbb{1}_n & \end{array} \right). \nonumber
\end{equation}
$\mathcal{J}$ and $J$ from \eqref{eq:TDualityGroupFerm} are connected via a similarity transformation
\begin{align*}
J = O_2^T O_1^T \mathcal{J} O_1 O_2 \quad \text{with} \ O_1 =\left( \begin{array}{c|c} \frac{1}{\sqrt{2}} \begin{scriptsize} \left( \begin{array}{cc} \mathbb{1}_{m} & \mathbb{1}_{m} \\  \mathbb{1}_{m} & -\mathbb{1}_{m}\end{array} \right) \end{scriptsize} & \\ \hline  & \mathbb{1}_{2n}\end{array} \right) \ \text{and} \ O_2 = \scriptsize{\left( \begin{array}{cccc} \mathbb{1}_m & & & \\ & 0_m & \mathbb{1}_{m\times n} & \\ & \mathbb{1}_{n\times m} & 0_n & \\ & & & \mathbb{1}_n \end{array} \right)}.
\end{align*}
} of the orthosymplectic group OSp$(d_b,d_b|2d_f)$ and nicely generalises the O$(d_b,d_b)$ group of bosonic $T$ duality. This group was previously introduced in \cite{Siegel:1993th}, see also \cite{Fre.Grassi.Sommovigo.Trigiante}. We will constrain further discussion of OSp$(d_b,d_b|2 d_f)$ to the generalisation of generic $TsT$ transformations \eqref{eq:TsT_SO(d,d)} of the bosonic case.

\subsubsection*{Fermionic Generalisation of $TsT$ Transformations}
Although along a fermionic coordinate $g_{T}^2 \neq \mathbb{1}$, the structure of the superduality group \eqref{eq:TDualityGroupFerm} does not become more complicated, since as mentioned above $T_\alpha^2$ is only a coordinate transformation $\theta_\alpha \rightarrow - \theta_{\alpha}$. As such we expect some fermionic analogue of the generic $TsT$ transformation \eqref{eq:TsT_SO(d,d)} to exist. For this we consider the \eqref{eq:TsT_SO(d,d)}-like ansatz
\begin{equation}
g_\Gamma = \left( \begin{array}{cc} \mathbb{1}_d &  \\ \Gamma & \mathbb{1}_d \end{array} \right)\, \label{eq:TsTinvFerm}.
\end{equation}
This lies in our representation \eqref{eq:TDualityGroupFerm} of OSp$(d_b,d_b|2 d_f)$ for
\begin{equation}
\Gamma = \left( \begin{array}{cc} \Lambda_b & \Omega \\ -\Omega^T & \Lambda_f \end{array} \right) \nonumber
\end{equation}
with a real skewsymmetric $d_b \times d_b$ matrix $\Lambda_b$, a Grassmann-valued $d_b \times d_f$ matrix $\Omega$ and a real symmetric $d_f\times d_f$ matrix $\Lambda_f$. Similarly to the bosonic case above, group elements of this type form an abelian subgroup of OSp$(d_b,d_b|2d_f)$.

The group element \eqref{eq:TsTinvFerm} now corresponds to a sequence of $Ts(T^{-1})$ transformations, with shifts defined as in \eqref{eq:shift_TsT}. Purely fermionic $Ts(T^{-1})$ transformations look like
\begin{equation}
g_{\Gamma_{f_1 f_2}} = g_{T_{f_1}} \cdot \left( \begin{array}{cccc} 1 & \gamma & & \\ 0 & 1 & & \\ & & 1 & 0 \\ & & -\gamma & 1 \end{array} \right) \cdot  g^{-1}_{T_{f_1}} =  \left( \begin{array}{cccc} 1 &  & & \\ & 1 & & \\ 0 & \gamma & 1 & \\ \gamma & 0 & & 1 \end{array} \right) \label{eq:singleTsTinvFerm}
\end{equation}
and indeed schematically $T_f s_{ff} T_f^{-1}$ give rise to symmetric, but off-diagonal entries in $\Lambda_f$ in \eqref{eq:TsTinvFerm}. It turns out that the diagonal elements in $\Lambda_f$ cannot be understood as a type $g_T \cdot g_{GL} \cdot g_T^{-1}$ transformation.\footnote{Up to $T$ duality transformations, the effect of diagonal elements of $\Lambda_f$ on the background is equivalent to a shift of $\mathcal{E}$. Namely $g_{\Lambda_{f,diag}} = T^{-1} \circ \left( \mathcal{E}_{\alpha \alpha} \rightarrow \mathcal{E}_{\alpha \alpha} + \Lambda_{f,\alpha \alpha} \right)\circ T\,, \, \quad \alpha=1,...,d_f .$} From here on, we therefore understand generic $Ts(T^{-1})$ transformations as group elements of OSp$(d_b,d_b|2d_f)$ of the type \eqref{eq:TsTinvFerm} with generic symmetric, but off-diagonal $\Lambda_f$.

Let us note that there is no ambiguity for $Ts(T^{-1})$ transformations ``mixing'' bosons and fermions: $T_f s_{fb} T^{-1}_f$- and $T_b s_{bf} T_b$-type transformations are equivalent and both correspond to the (skewsymmetric) odd part of $\Gamma$ in \eqref{eq:TsTinvFerm}. Of course $Ts(T^{-1})$ transformations directly reduce to $TsT$ transformations if the $T$ duality is a bosonic one and so, for the sake of simplicity, we will refer to $Ts(T^{-1})$ transformations as $TsT$ transformations from now on. Both only differ by a trivial coordinate redefinition in any case.

\section{Equivalence of Abelian Yang-Baxter Deformations and $TsT$ Transformations}
\label{chap:Supergravity}

In this section we prove that any Yang-Baxter deformation generated by an abelian solution to the CYBE is equivalent to a $TsT$ transformation at the level of the corresponding $\sigma$ model.

This equivalence was previously proposed in \cite{Tongeren.CYBE-deformations}, and is supported by many examples, see e.g. \cite{Matsumoto:2014nra,Matsumoto:2014gwa,MatsumotoYoshida.Gravity-CYBE}, but a general proof is still missing. We will also extend this claim by considering $r$ matrices built out of anticommuting supercharges. Using a parameterisation of the coset manifold with manifest shift invariance in $d = d_b + d_f$ coordinates, we will prove that the (coordinate-dependent) $TsT$ transformation behaviour \eqref{eq:TsTinvFerm} can be reproduced by an abelian $R$ operator, and vice versa. As the Yang-Baxter deformed action \eqref{eq:DeformedLagCYBE} is independent of parameterisation this introduces a coordinate-independent notion of $TsT$ transformations in the form of abelian Yang-Baxter deformations.

\subsection{Natural Parameterisation with Manifest Shift Isometries}
The starting point of our proof is to choose a natural parameterisation of the coset manifold where we have shift isometries in the coordinates associated to (anti)commuting generators $t_a$, namely
\begin{equation}
g = \exp(Z^a t_a)\bar{g}(Z^{\underline{a}}). \label{eq:Parameterisation}
\end{equation}
There the $Z^a$ are the $d=d_b + d_f$ isometry coordinates and $Z^{\underline{a}}$ are the remaining coordinates, $Z^M = (Z^a,Z^{\underline{a}}) = (X^i,\theta^\alpha,Z^{\underline{a}})$. $\bar{g}$ is assumed to be chosen in a way that the metric is non-degenerate, so we can consider \eqref{eq:Parameterisation} to be a valid parameterisation of the coset manifold. This is motivated for instance by the group parameterisations of AdS$_N$ in Poincaré coordinates as
\begin{equation}
g_{AdS} = e^{X^\mu p_\mu} z^{ -D } \nonumber, \qquad \text{with} \ \mu=0,1,2,...,N-2
\end{equation}
where $p^\mu$ respectively $D$ are the momentum respectively dilatation generators of the conformal algebra $\mathfrak{so}(2,N-1)$. There we have $N-1$ isometries parameterised by $X^\mu$, as $[ p^\mu , p^\nu ] = 0 $ by means of the conformal algebra. This type of group parameterisation should always be possible for general group and coset manifolds and any choice of (anti)commuting generators $t_a$ in the symmetry algebra. Let us sketch a proof for the bosonic case.

We assume that we have a geometry with $d$ commuting Killing vector fields. Then there are coordinates $Z^M = (X^i, Y^{\underline{i}})$ in which these vector fields are $\frac{\partial}{\partial X^i}$, thus the commuting isometries are parameterised by $X^i$. In particular, the background and a choice of a local frame ${e_\mu}^a$ with a corresponding spin connection ${\omega_\mu}^{ab}$ are independent of the $X^i$.

The Maurer-Cartan form on a coset manifold (see e.g. \cite{Metsaev.Tseytlin.TypeIIbGSAction.AdS5S5}) decomposes into
\begin{equation}
A = -g^{-1} \mathrm{d} g = {e_\mu}^a P_a \mathrm{d} X^\mu + {\omega_\mu}^{ab} J_{ab} \mathrm{d} X^\mu
\end{equation}
with coset generators $P_a$ and isotropy generators $J_{ab}$, so in our case
\begin{equation}
A = A_i(Y) \mathrm{d} X^i + A_{\underline{i}}(Y) \mathrm{d} Y^{\underline{i}}. \nonumber
\end{equation}
The flatness of $A$ implies that
\begin{align*}
[A_i (Y) , A_j (Y) ] = 0 \quad \text{due to} \ \partial_i A_j = 0 \qquad \forall i,j = 1,...,d .
\end{align*}
For every $Y$ these span a $d$-dimensional commuting algebra. It follows there is similarity transformation with a group valued function $g_2(Y)$
\begin{align}
A_i (Y) = g_2^{-1} (Y) h_i g_2(Y) \quad \forall i = 1,...,d \ , \label{eq:Parameterisation2}
\end{align}
where the $h_i$ are the constant commuting generators of the algebra corresponding to the Lie algebra of the commuting Killing vector fields.\footnote{In the non-compact case there are inequivalent choices of commuting subalgebras/isometries. These inequivalent choices would correspond to different choices of our Killing vector fields at the beginning of the proof.} Note that we use the notation $h_i$ for a general set of commuting generators, which in the non-compact case will generically not be the Cartan generators.

Now consider a group parameterisation $\tilde{g} = \exp(X^i h_i) g_2(Y)$ with $\tilde{A} = - \tilde{g}^{-1} \mathrm{d} \tilde{g}$. It follows that
\begin{align*}
\tilde{A}_i &= A_i \quad \Rightarrow \quad g = g_1(Y)  \exp(X^i h_i) g_2(Y) \qquad \text{for some} \ g_1(Y) .
\end{align*}
Again from the flatness of $A$ follows that
\begin{align*}
\partial_i A_{\underline{j}} = \partial_{\underline{j}} A_i + [A_i, A_{\underline{j}}] &= 0 \quad \Rightarrow \quad [A_i, A_{\underline{j}}] = [A_i, \tilde{A}_{\underline{j}}] \\
\Rightarrow \quad [ \text{Ad}_{\tilde{g}}^{-1}(- g_1^{-1} \partial_{\underline{j}} g_1) , A_i ] &= \text{Ad}_{\tilde{g}}^{-1} \left( [- g_1^{-1} \partial_{\underline{j}} g_1 , h_i ] \right) = 0,
\end{align*}
so that $g_1$ is generated by the $h_i$. It follows that a group parameterisation of the form
\begin{align}
g = \exp(X^i h_i) g_1(Y) g_2(Y) \equiv \exp(X^i h_i) \bar{g}(Y) \label{eq:Parameterisation3}
\end{align}
exists for any choice of commuting generators $h_i$.

\subsection{Bosonic Abelian Yang-Baxter Deformations}
Now consider a generic abelian $r$ matrix that consists some bosonic commuting generators $h_i$ of the global symmetry algebra of the coset model
\begin{equation}
r = - \tilde{\Gamma}^{ij} h_i \wedge h_j, \label{eq:rAbelian}
\end{equation}
with a (real) antisymmetric $d \times d$ parameter matrix $\tilde{\Gamma}^{ij}$.
Consider a parameterisation of the form \eqref{eq:Parameterisation},
\begin{equation}
g = \exp(X^i h_i)\bar{g}(Y).
\end{equation}
Due to the fact that the $h_i$ commute, the Maurer-Cartan form becomes
\begin{equation}
A = - g^{-1} \mathrm{d} g = - \text{Ad}_{\bar{g}}^{-1} ( \mathrm{d} X^i h_i ) + \bar{A}(Y) = - \text{Ad}_{g}^{-1} ( h_i ) \mathrm{d} X^i + \bar{A}(Y) \equiv A_i(Y) \mathrm{d} X^i + \bar{A}(Y) ,
\end{equation}
and the Lagrangian is manifestly shift-invariant in the $X^i$-coordinates. With this we see that the abelian $r$ matrix \eqref{eq:rAbelian} is actually built from some components of the conserved currents with respect to the global symmetry of the coset $\sigma$ model, $A^R = \text{Ad}_g (A) = - \mathrm{d}g \ g^{-1}$ . The corresponding dressed $r$ matrix then is
\begin{equation}
r_g = \left(\text{Ad}_{g}^{-1} \otimes \text{Ad}_{g}^{-1} \right) \cdot r
\end{equation}
and the associated linear $R$ operator can be expressed nicely in terms of the Maurer-Cartan form components
\begin{equation}
r_g = - \tilde{\Gamma}^{ij} A_i \wedge A_j \quad \Rightarrow \quad R_g( M ) = \text{STr}_2 \left( r_g \cdot ( \mathbb{1} \otimes M ) \right) = - \tilde{\Gamma}^{ij} A_i \text{STr}( A_j M ). \label{eq:rMatrixMoreGeneral}
\end{equation}
Writing
\begin{equation}
\Gamma = \left( \begin{array}{cc} \tilde{\Gamma} & \\ & 0_{D-d} \end{array} \right), \nonumber
\end{equation}
it follows that
\begin{align*}
 R_g \circ d_- (A_N) &= - \tilde{\Gamma}^{ij} A_i \text{STr}\left(A_j d_-(A_M) \right) = A_M {(-\Gamma \mathcal{E})^M}_N \\
( R_g \circ d_-)^n (A_N) &=  A_M {((-\Gamma \mathcal{E})^n)^M}_N.
\end{align*}
The Yang-Baxter deformed Lagrangian \eqref{eq:DeformedLagCYBE} then becomes
\begin{equation}
\mathcal{L} \propto \partial_+ X^M \tilde{\mathcal{E}}_{MN} \partial_- X^N
\end{equation}
with the general coordinates $X^M = (X^i , Y^{\underline{i}})$ and the deformed background
\begin{align}
\tilde{\mathcal{E}}_{MN} &= \text{STr} \left( A_M d_- \circ \frac{1}{1 - R_g \circ d_-} (A_N) \right) \nonumber \\
&= \sum_{n=0}^\infty \text{STr} \left( A_M d_- \circ (R_g \circ d_-)^n (A_N) \right) = \sum_{n=0}^\infty \text{STr} \left( A_M d_- (A_K) \right){((-\Gamma \mathcal{E})^n)^K}_N \nonumber \\
&= \mathcal{E}_{MK} {\left(( \mathbb{1} + \Gamma \mathcal{E})^{-1}\right)^K}_N. \label{eq:deformed_background}
\end{align}
This directly corresponds to the O$(d,d)$ group element  \eqref{eq:TsT_SO(d,d)} describing a generic bosonic $TsT$ transformation.

\subsection{Inclusion of Fermions}
A generic abelian graded skewsymmetric $r$ matrix over a Lie superalgebra in our conventions is built out of (anti)commuting even (odd) generators $ \{ h_i , Q_\alpha \} $ with
\begin{align*}
[h_i,h_j] = 0, \qquad [h_i,Q_\alpha] = 0 \qquad \{ Q_\alpha , Q_\beta \} = 0 \quad \text{for $i,j=1,...,d_b$ and  $\alpha,\beta = 1,...,d_f$},
\end{align*}
as
\begin{equation}
r = - \Lambda_b^{ij} h_i \wedge h_j - \Omega^{i\alpha} h_i \wedge
Q_\alpha - \Omega^{\alpha i} Q_\alpha \wedge
h_i - \Lambda_f^{\alpha \beta} Q_\alpha \wedge Q_\beta \equiv - \tilde{\Gamma}^{ab} t_a \wedge t_b, \label{eq:rAbelianFerm}
\end{equation}
with $t_a = (h_i , Q_\alpha)$ and a graded skewsymmetric $(d_b|d_f) \times (d_b|d_f)$-matrix
\begin{align*}
\tilde{\Gamma} = \left( \begin{array}{cc} \Lambda_b & \Omega \\ -\Omega^T &  \Lambda_f \end{array} \right) .
\end{align*}
Here $\Lambda_f$ is a symmetric, but off-diagonal real $d_f \times d_f$-matrix, $\Omega$ is an arbitrary Grassmann-valued $d_b \times d_f$-matrix and $\Lambda_b$ is a skewsymmetric real $d_b \times d_b$-matrix. We should emphasize that $\mathfrak{su}(2,2|4)$ and $\mathfrak{psu}(2,2|4)$ do not contain real supercharges that anticommute with themselves, so these fermionic extensions of abelian $r$ matrices do not exist for the real $\ads$ superstring, or its AdS$_3$ and AdS$_2$ cousins. To consider them we need to work with the complexified model. The $r$ matrices are then complex and break reality of the action, but are otherwise admissible.

With some care\footnote{This is rather tedious with our conventions, as for the fermionic Maurer-Cartan components
\begin{align*}
A^\Theta := A^r_\Delta \mathrm{d} \theta^\Delta = \mathrm{d} \theta^\Delta A^l_\Delta \qquad \text{with e.g.} \ A^r_\alpha = - g^{-1} Q_\alpha  (g^{ST})^{ST}) .
\end{align*}
It is important to pay attention to some subtleties of the graded tensor product in the definition of $r_g = ( \text{Ad}_g^{-1} \otimes \text{Ad}_g^{-1}) \cdot r$ which match the above ambiguity and lead to the desired $R_g$ operator in \eqref{eq:FermCalc}.} regarding the Grassmann-valued fields $\theta$ the proof works in the same way as in the bosonic case. First we choose a group parameterisation with manifest isometries corresponding to the (anti)commuting generators and express the $R_g$ operator corresponding to \eqref{eq:rAbelianFerm} by some components of the Maurer-Cartan form.
\begin{align}
g &= \exp(X^i h_i + \theta^\alpha Q_\alpha) \bar{g}(Z^{\underline{a}}) \\
A &= - \text{Ad}_g^{-1} (\mathrm{d}X^i h_i + \mathrm{d} \theta^\alpha Q_\alpha) + \bar{A}(Z^{\underline{a}}) \nonumber \\
&\equiv - A_i \mathrm{d} X^i - A_\alpha^r \mathrm{d} \theta^\alpha + \bar{A} (Z^{\underline{a}}) = - A_i \mathrm{d} X^i - \mathrm{d} \theta^\alpha A_\alpha^l + \bar{A} (Z^{\underline{a}}) \nonumber \\
R_g(M) &= - A^r_a \tilde{\Gamma}^{a b} \text{STr}(A^l_b M) \label{eq:FermCalc}
\end{align}
The undeformed background $\mathcal{E}_{MN}$ is given terms of the components of the Maurer-Cartan form in the conventions of \eqref{eq:ActionSigma} and \eqref{eq:UndeformedLag} by
\begin{align*}
\mathcal{E}_{MN} = \text{STr}(A_M^l \ d_- (A_N^r) ),
\end{align*}
so we get $(R_g \circ d_-)^n (A^r_N) = A_M^l ({(- \Gamma \mathcal{E})^n )^M}_N$ with $\Gamma = \left( \begin{array}{cc} \tilde{\Gamma} & \\ & 0_{D-d_b-d_f} \end{array} \right)$.

In the same way as in the bosonic case the abelian Yang-Baxter deformation results in a deformed background
\begin{align*}
\tilde{\mathcal{E}} = \mathcal{E}(\mathbb{1} + \Gamma \mathcal{E})^{-1}.
\end{align*}
In other words, we directly reproduce the generic $TsT$ transformation behaviour \eqref{eq:TsTinvFerm} of the superduality group OSp$(d_b,d_b|2d_f)$, and vice versa.

The direct approach via a natural parameterisation with manifest isometries like \eqref{eq:Parameterisation} is useful to see the $TsT$ behaviour of abelian Yang-Baxter deformations as in \eqref{eq:TsT_SO(d,d)}, in particular to determine its effect on the concrete background. The abelian Yang-Baxter deformation in the form \eqref{eq:DeformedLagCYBE} on the other hand, gives a coordinate-independent representation of $TsT$ transformations (in contrast to the OSp$(d_b,d_b|2d_f)$-approach).
Moreover this manifestly shows that every $TsT$ transformation of such a (super)coset gives an integrable model with \eqref{eq:DeformedLaxCYBE} as the associated Lax pair.

Abelian Yang-Baxter deformed models correspond to supergravity solutions by construction, as $T$ duality and thus $TsT$ transformations map two supergravity solutions to each other \cite{Bergshoeff.Hull.Ortin.TDuality}, also in the fermionic case \cite{Berkovits.Maldacena.fermT-Duality}.\footnote{In terms of the action on the background fields, the standard treatment of $T$ duality for a supergravity background coupling to a Green-Schwarz superstring \cite{Kulik.Roiban.TDuality,Hassan.TDualityCurvedBackgrounds} does not admit an immediate O$(d,d)$-like formulation of $TsT$ transformations. However, an appropriate extension to the Ramond-Ramond forms exists \cite{Fukuma.Oota.Tanaka.T-Duality.RR-Pot,Brace.Morariu.Zumino.TDuality.RR-background.Matrix.Model,Hassan.SOdd.Spinors}. The action of the superduality group OSp$(d_b,d_b|2d_f)$ on the supergravity fields has not been investigated yet to our knowledge. For fermionic $T$ duality transformations themselves some progress was made in \cite{Sfetsos.Siampos.Thompson.FermionicTDuality} in the canonical formulation. $TsT$ transformations including fermions were studied previously in \cite{Alday:2005ww} for deformations of S$^5$ in the $\sigma$ model approach.} This matches the analysis of \cite{Borsato.Wulff2016}, as any abelian $r$ matrix is unimodular.


\section{On Inequivalent $TsT$ Transformations}
\label{chap:AdS}
In this section we want to illustrate the fact that there are  different inequivalent sets of commuting shift isometries and thus $TsT$ transformations on non-compact backgrounds. For completeness we start with $TsT$ transformation of S$^3$.

\subsection{Sphere S$^3$}

We have seen in the previous section that a natural parameterisation of the background with $d$ commuting isometries is $g = \exp(X^i h_i) \  \bar{g}$ with a choice of $d$ commuting generators $\{ h_i \}$. As S$^N$ and its isometry group O$(N+1)$ is compact, any other choice of the commuting generators $\{ k_i\}$ is connected via a similarity transformation with a group element $S$ related to the $\{ h_i \}$ as $k_i = S h_i S^{-1}$. Exactly as in \eqref{eq:Parameterisation2} the corresponding group element
\begin{equation}
g_k = \exp(X^i k_i) \ S \bar{g} \quad \Rightarrow \quad A_k = -g_k^{-1} \mathrm{d} g_k = A
\end{equation}
yields the same background as $g$ because $S$ is constant.

We work with generators $n _{ij}$ of $\mathfrak{so}(N+1)$, satisfying
\begin{align*}
[ n_{ij} , n_{kl} ] = \delta_{il} n_{jk} - \delta_{jl} n_{ik} - \delta_{ik} n_{jl} + \delta_{jk} n_{il} \qquad i,j,k,l = 1,...,N+1.
\end{align*}

S$^3$ is the minimal example for the study of $TsT$ transformations on spheres, with the rank of $\mathfrak{so}(4)$ being two. We choose $n_{12}, n_{34}$ as the Cartan basis, $r = - \gamma \ n_{12} \wedge n_{34}$ and the corresponding group parameterisation with manifest isometries to be
\begin{align}
\exp\left(\phi_1 n_{12} +  \phi_2 n_{34} \right) \exp(\theta n_{24}).
\end{align}
This corresponds to the metric
\begin{align}
(\mathrm{d} s)^2 &= \sin^2 \theta (\mathrm{d} \phi_1)^2 + \cos^2 \theta (\mathrm{d} \phi_2)^2 + \left( \mathrm{d} \theta \right)^2. \nonumber
\end{align}
The $TsT$ deformed three-sphere looks like
\begin{align}
(\mathrm{d} s)^2_{def} &= \frac{1}{1 + \frac{\gamma^2}{8}(1 - \cos(4 \theta))}\left(\sin^2 \theta (\mathrm{d} \phi_1)^2 + \cos^2\theta (\mathrm{d} \phi_2)^2\right) + \left( \mathrm{d} \theta \right)^2 \nonumber \\
B_{def} &= \frac{\frac{\gamma}{2} \sin^2(2 \theta) }{1 + \frac{\gamma^2}{8}(1 - \cos(4 \theta))} \mathrm{d} \phi_1 \wedge \mathrm{d} \phi_2. \label{eq:S3Deformed}
\end{align}

\subsection{Anti-de Sitter Space AdS$_3$}

In the non-compact case there are inequivalent choices of commuting generators. We will only explicitly discuss the inequivalent deformations of AdS$_3$, where this undertaking is greatly simplified due to the structure of $\mathfrak{so}(2,2)$. This gives some insight in the various possible abelian Yang-Baxter deformations of AdS$_5$.

The symmetry algebra of AdS$_3$ is $\mathfrak{so}(2,2)$, which has the nice decomposition\footnote{This structure essentially makes it possible to independently deform the two factors also for quantum deformations \cite{Hoare:2014oua}.}
\begin{equation}
\mathfrak{so}(2,2) \simeq \mathfrak{sl}(2,\mathbb{R}) \oplus  \mathfrak{sl}(2,\mathbb{R}).
\end{equation}
From here we can immediately read off all possible commuting isometries, namely one arbitrary element of each factor. We work with the following representation of $ \mathfrak{sl}(2,\mathbb{R}) $
\begin{align*}
 h = \left( \begin{array}{cc} 1 & 0 \\ 0 & -1 \end{array} \right), \ a &= \left( \begin{array}{cc} 0 & 1 \\ 0 & 0 \end{array} \right), \ b = \left( \begin{array}{cc} 0 & 0 \\ 1 & 0 \end{array} \right) \\
[ h , a ] = 2a \ , \quad [h,b] &= - 2 b \ , \quad [a,b] = h
\end{align*}
and $\mathfrak{so}(2,2)$ generators $m_{ij}$ resp. conformal generators $p_\mu,k_\mu,D,m_{01}$
\begin{align*}
[ m_{ij} , m_{kl} ] &= \eta_{il} m_{jk} - \eta_{jl} m_{ik} - \eta_{ik} m_{jl} + \eta_{jk} m_{il} \qquad i,j,k,l = 0,...,3 \\
\eta &= \text{diag}(-1,1,1,-1) \\
p_\mu &= m_{\mu 2} + m_{\mu 3}, \quad k_\mu = m_{\mu 2} - m_{\mu 3} \quad \text{and} \quad D= m_{23} \qquad \mu = 1,2.
\end{align*}
Then we see that the two copies of $ \mathfrak{sl}(2,\mathbb{R}) $ in  $\mathfrak{so}(2,2)$ are spanned by
\begin{align*}
h_1 = m_{01} - D \qquad a_1 = p_+ \qquad b_1 = k_-
\end{align*}
respectively
\begin{align*}
h_2 = m_{01} + D \qquad a_2 = k_+ \qquad b_2 = p_-
\end{align*}
with $v_\pm := \frac{1}{2}(v_0 \pm v_1)$. Explicitly, generic abelian $r$ matrices are of the form
\begin{align}
r = s_1 \wedge s_2 \qquad \text{with} \qquad (s_1 , s_2) \in \mathfrak{sl}(2,\mathbb{R}) \oplus \mathfrak{sl}(2,\mathbb{R}) \simeq \mathfrak{so}(2,2). \label{eq:rMatrixGenso(22)}
\end{align}

From the point of view of the Yang-Baxter deformations the overall scaling of the $r$ matrix only contributes to the deformation parameter, so for each factor in \eqref{eq:rMatrixGenso(22)} we only need to consider $\det s < 0$, $\det s >0$ or $\det s = 0 $. These three classes of generators are clearly inequivalent to each other under similarity transformations $\tilde{s} = S s S^{-1}$ with $S \in \text{SL}(2,\mathbb{R})$. $\text{SL}(2,\mathbb{R})$ moreover acts transitively on each class (up to rescaling). Convenient representants are
\begin{enumerate}
\item $\det s = -1$: $s \sim h$
\item $\det s = 0$: $s \sim a$
\item $\det s = 1$: $s \sim a-b$.
\end{enumerate}
We can now combine these $\mathfrak{sl}(2,\mathbb{R})$ generators of both copies in $\mathfrak{so}(2,2)$ to a generic $r$ matrix. Exchanging the two copies of $ \mathfrak{sl}(2,\mathbb{R})$ is an outer automorphism of $\mathfrak{so}(2,2)$
\begin{align*}
h_1 \leftrightarrow h_2 \qquad a_1 \leftrightarrow a_2 \qquad b_1 \leftrightarrow b_2
\end{align*}
The physical interpretation is either
\begin{equation}
D \leftrightarrow -D, \quad p \leftrightarrow k \qquad \text{or} \qquad D \leftrightarrow -D, \quad + \leftrightarrow -. \label{eq:AdSAuto1}
\end{equation}

With use of \eqref{eq:AdSAuto1} we are left with six types of abelian $r$ matrices, namely:
\begin{itemize}
\item $h_1 \wedge h_2$ corresponds to the (non-compact) Cartan $r$ matrix  $r = -\gamma m_{01}\wedge D$. A convenient parameterisation is given by $g=\exp\left(\theta m_{01} + \ln (z) D \right) \exp((u z) p_0)$, corresponding to the metric
\begin{align*}
(\mathrm{d} s)^2 = - (z \mathrm{d} u)^2 + (uz)^2 (\mathrm{d} \theta)^2 + \left( \mathrm{d} \ln(z) \right)^2
\end{align*}
of hyperpolar Poincaré coordinates. A coordinate change $u \rightarrow x/z$ yields  $\ln(z)$ and the boost-angle $\theta$ as isometry coordinates. The associated Yang-Baxter deformed background reads
\begin{align}
(\mathrm{d} s)^2_{def} &= \frac{1}{ 1 + \gamma^2 (u z)^2- \gamma^2 (u z)^4 } \left( - ( 1 + \gamma^2 (u z)^2) z^2 (\mathrm{d} u)^2 + (u z)^2 (\mathrm{d} \theta)^2  \right. \nonumber \\
&{} \qquad \left. - 2 \gamma^2 u^3 z^4 \ \mathrm{d} u \ \mathrm{d} \ln(z) + \left(  1 - \gamma^2 (u z)^4  \right) \left( \mathrm{d} \ln(z) \right)^2 \right), \nonumber \\
B_{def} &= \frac{ 2 \gamma (u z)^2 ( z^2 u \ \mathrm{d} u + \mathrm{d} \ln(z)) }{1 + \gamma^2 (u z)^2- \gamma^2 (u z)^4 } \wedge \mathrm{d} \theta, \label{eq:DeformHH}
\end{align}
in terms of the original hyperpolar Poincar\'e coordinates.

\item $(a_1 - b_1) \wedge ( a_2 - b_2 )$ translates to the (compact) Cartan $r$ matrix $r = - \gamma \ m_{03} \wedge m_{12}$ leading to a $TsT$ transformation corresponding to time shifts and spatial rotations. These are natural in global coordinates, where both isometries are manifest. With a group parameterisation $g = \exp( \phi m_{03} + \theta m_{12} ) \exp (\rho m_{23} )$ the undeformed and deformed  backgrounds are
\begin{align}
(\mathrm{d} s)^2 &= - \cosh^2 \rho (\mathrm{d} \phi)^2 + \sinh^2 \rho (\mathrm{d} \theta)^2 + \left( \mathrm{d} \rho \right)^2 ,\nonumber \\
(\mathrm{d} s)^2_{def} &= \frac{1}{1 + \frac{\gamma^2}{8}(1 - \cosh(4 \rho))}\left(- \cosh^2 \rho (\mathrm{d} \phi)^2 + \sinh^2 \rho (\mathrm{d} \theta)^2\right) + \left( \mathrm{d} \rho \right)^2 ,\nonumber \\
B_{def} &= \frac{\frac{\gamma}{2} \sinh^2(2 \rho) }{1 + \frac{\gamma^2}{8}(1 - \cosh(4 \rho))} \mathrm{d} \phi \wedge \mathrm{d} \theta .\label{eq:DeformHHglobalcoo}
\end{align}

\item $a_1 \wedge a_2$ corresponds to $\tilde{r} = -\gamma p_+ \wedge p_- \propto r = -\gamma p_0 \wedge p_1 $. With group parameterisation $g = \exp(-x_0 p_0 + x_1 p_1) \ z^D$ the undeformed and deformed backgrounds are
\begin{align}
(\mathrm{d} s)^2 &= z^2 \left( - (\mathrm{d} x_0)^2 + (\mathrm{d} x_1)^2 \right) + \left( \mathrm{d} \ln(z) \right)^2, \nonumber \\
(\mathrm{d} s)^2_{def} &= \frac{z^2}{1 - \gamma^2 z^4} \left( - (\mathrm{d} x_0)^2 + (\mathrm{d} x_1)^2 \right) + \left( \mathrm{d} \ln(z) \right)^2  ,\nonumber \\
B_{def} &= \frac{2 \gamma z^4}{1 - \Gamma^2 z^4} \mathrm{d} x_0 \wedge \mathrm{d} x_1 .\label{eq:DeformPPpoincare}
\end{align}
\end{itemize}
The manifest isometry coordinates for the remaining three $r$ matrices are not very intuitive as the $r$ matrices mix the generators corresponding to costumary choices of coordinates (like global or Poincaré coordinates). We therefore give the deformed backgrounds in light-cone Poincaré coordinates (group parameterisation $g = \exp(x_+ p_- + x_- p_+) \ z^D$)
\begin{align*}
(\mathrm{d} s)_{undef}^2 &= - z^2 \mathrm{d} x_+ \mathrm{d} x_- + \left( \mathrm{d} \ln(z) \right)^2 \nonumber.
\end{align*}
\begin{itemize}
\item $h_1 \wedge a_2$: $r = - \gamma (m_{01} - D)\wedge p_-$
\begin{align}
(\mathrm{d} s)^2_{def} &= - C \left( \frac{\gamma^2}{4} z^4 \ (\mathrm{d} x_-)^2 +  z^2 \ \mathrm{d} x_+ \mathrm{d} x_- + \gamma^2 x_- z^3 \ \mathrm{d} z \mathrm{d} x_- \right) + \left( \mathrm{d} \ln(z) \right)^2 , \nonumber \\
B_{def} &= \gamma \ C \left( x_- \ z^4 \ \mathrm{d} x_- \wedge \mathrm{d} x_+ + z \ \mathrm{d} x_- \wedge \mathrm{d} z \right) .\label{eq:DeformHPpoincareLC}
\end{align}
with $C^{-1} = 1 - \gamma^2 x_-^2 z^4$.
\item $h_1 \wedge (a_2 - b_2)$: $r = - \gamma (m_{01} - D) \wedge (p_- - k_+)$
\begin{align}
(\mathrm{d} s)^2_{def} &= - C \left( \frac{\gamma^2}{4}(1 + x_+^2) ^2 z^4 \ (\mathrm{d} x_-)^2 +  z^2 \left( 1- \frac{\gamma^2}{2} (2 x_- x_+ (1 + x_+^2) z^2 - x_+^2 - 1) \right) \ \mathrm{d} x_- \mathrm{d} x_+  \right. \nonumber \\
&{} \quad + \gamma^2 x_- (1+x_+^2)^2 z^3 \ \mathrm{d} x_- \mathrm{d} z + \frac{\gamma}{4} (1 - 2 x_- x_+ z^2)^2 \ (\mathrm{d} x_+)^2 \nonumber \\
&{} \quad - \left. \gamma^2 x_- (1+x_+^2) z (1-2x_- x_+ z^2) \ \mathrm{d} x_+ \mathrm{d} z - \frac{1 - \gamma^2 x_-^2 (1 + x_+^2)^2 z^4}{z^2} \ (\mathrm{d} z)^2 \right) \nonumber ,\\
B_{def} &= - \gamma \ C \left( x_- \ (1 + x_+^2) \ z^4 \ \mathrm{d} x_- \wedge \mathrm{d} x_+ + (1 + x_+^2) z \ \mathrm{d} x_- \wedge \mathrm{d} z + (1 - 2x_- x_+ z^2) \ \mathrm{d} x_+ \wedge \mathrm{d} z \right). \label{eq:DeformHComppoincareLC}
\end{align}
with $C^{-1} = 1 - \gamma^2 \left( 1 + (x_+ - x_-(1 + x_+^2) z^2)^2 \right) $.
\item $(a_1 - b_1) \wedge a_2$: $r = - \gamma (p_+ - k_-) \wedge p_- $
\begin{align}
(\mathrm{d} s)^2_{def} &= - C \left( \frac{\gamma^2}{4} x_-^2 z^4 \ (\mathrm{d} x_-)^2 +  z^2 \ \mathrm{d} x_+ \mathrm{d} x_- + \frac{\gamma^2}{2} x_- (1 + x_-^2) z^3 \ \mathrm{d} z \mathrm{d} x_- \right) + \left( \mathrm{d} \ln(z) \right)^2 , \nonumber \\
B_{def} &= - \frac{1}{2} \gamma \ C \left( (1+x_-^2) \ z^4 \ \mathrm{d} x_- \wedge \mathrm{d} x_+ + x_- \ z \ \mathrm{d} x_- \wedge \mathrm{d} z \right) . \label{eq:DeformCompPpoincareLC}
\end{align}
with $C^{-1} = 1 - \frac{\gamma^2}{4} (1 + x_-^2)^2 z^4$.
\end{itemize}

\subsubsection*{AdS$_5$}
The conformal symmetry of AdS$_5$ does not decompose nicely as in the AdS$_3$ case, and we will not give an extensive list of inequivalent $TsT$ transformations here. To illustrate the extent of the full list, note that we could for instance consider abelian Yang-Baxter deformations based on the subalgebras
\begin{align}
\mathfrak{so}(2,4) &\supset \mathfrak{so}(2,2) \oplus \mathfrak{so}(2)_{space} \simeq \mathfrak{sl}(2,\mathbb{R}) \oplus \mathfrak{sl}(2,\mathbb{R}) \oplus \mathfrak{so}(2)_{space}, \nonumber \\
\mathfrak{so}(2,4) &\supset \mathfrak{so}(2)_{time} \oplus \mathfrak{so}(4) \simeq \mathfrak{so}(2)_{time} \oplus \mathfrak{su}(2) \oplus \mathfrak{su}(2) ,\nonumber \\
\text{or} \qquad \mathfrak{so}(2,4) &\simeq \mathfrak{conf}(1,3) \supset \text{span}(p_\mu) \quad \text{or} \quad \text{span}(k_\mu) \label{eq:decomposition},
\end{align}
leading to many tens of inequivalent deformations already. A method to obtain and classify all inequivalent commuting subalgebras of $\mathfrak{so}(2,4)$ and thus also abelian Yang-Baxter deformations was proposed in principle in \cite{Patera.Winternitz.Zassenhaus.ContinuousSubgroups}. In addition to pure AdS$_5$ deformations we could of course mix AdS$_5$ and S$^5$ directions.

\section{Conclusion and Outlook}

In this paper we proved that abelian Yang-Baxter deformations are equivalent to sequences of commuting $TsT$ transformations. This proof is completely generic and holds for any group or (semi-)symmetric coset $\sigma$ model, including fermions to all orders. We included the fermionic generalisation of these transformations, which however typically requires complexification. Including fermionic transformations naturally leads to a TsT subgroup of the superduality group OSp$(d_b,d_b|2d_f)$ generalising the bosonic $T$ duality group O$(d_b,d_b)$.

For illustrative purposes we moreover presented all six possible inequivalent abelian deformations of $\mathrm{AdS}_3$. In terms of the $\mathfrak{so}(2,2)$-generators the associated $r$ matrices are given by
\begin{align*}
m_{01} &\wedge D,& m_{03} &\wedge m_{12},& p_0 &\wedge p_1, \\
(m_{01}-D) &\wedge p_-,& (m_{01}-D) &\wedge (p_- - k_+) & (p_+ - k_-) &\wedge p_-.
\end{align*}

One natural question to ask is what the dual field theory interpretation of Yang-Baxter deformations is. For $r$ matrices solving the regular classical Yang-Baxter equation -- which includes the present abelian ones -- these duals are generically conjecture to be noncommutative versions of supersymmetric Yang-Mills theory \cite{vanTongeren:2015uha}, provided they exist. This conjecture relies on the twisted symmetry structure of the gravitational models, whose realisation on the hypothetical field theory side requires a nontrivial star product. Several abelian deformed theories are known to fit this description, notably the gravity duals of $\beta$ deformed SYM \cite{Lunin.Maldacena.Deformation} and canonical spacelike noncommutative SYM \cite{Hashimoto:1999ut,Maldacena:1999mh}. As discussed in \cite{vanTongeren:2015uha}, the situation is less clear for the naive time-like noncommutative version of SYM and the related abelian deformation of $\ads$ for example. The generalisation from the $\beta$ to the $\gamma_i$ deformation \cite{Frolov.LaxPair.Lunin-Maldacena.Background} shows subtleties as well, though at least in the spectrum a notion of duality appears to remain, see e.g. \cite{Fokken:2013aea,vanTongeren:2013gva,Fokken:2014soa}. It is important to understand in which (isolated) cases, and how, the general dual field theory picture breaks down.

In principle we can formally extend the conjecture of \cite{vanTongeren:2015uha} to our fermionic $TsT$  transformations, replacing field products in the SYM Lagrangian by star products built on the twist $e^{i \gamma r}$, where $r$ is associated $r$ matrix. As such $r$ matrices are not real, however, this would be a complex deformation of SYM. Moreover, manifest conformal invariance would be broken, cf. eqn. \eqref{eq:DeformedSymmetryGenerators}.\footnote{Suitably chosing an anticommuting supercharge $Q$ and superconformal $S$, it is possible to preserve scale invariance, at least classically. Fermionic abelian deformations always break Lorentz invariance however.} In particular such star products introduce new, possibly dimensionful, couplings in the theory. On the gravity side it would be useful to gain a better understanding of the action of fermionic $TsT$ transformations on the supergravity fields (and their reality). Duals of mixed bosonic-fermionic deformations could be defined similarly, though the nature of their deformation parameter is slightly odd.

There are a number of further open questions. First, it would be interesting to consider classical solutions and associated integrable classical mechanical models for these abelian deformed models, as well as non-abelian ones, as done for the $\beta$ deformation \cite{Frolov:2005ty}, and the $\eta$ model in e.g. \cite{Arutyunov:2014cda,Banerjee:2014bca,Kameyama:2014vma,Arutyunov:2016ysi,Banerjee:2016xbb}. Second, given the classical equivalence between the $\eta$ and $\lambda$ models via Poisson-Lie duality (cf. footnote \ref{footnote:lambdapoisson}), we might wonder whether similar dual theories exist for CYBE-based deformations. Third, non-Cartan abelian deformations (and non-abelian ones) invariably break the isometries required to fix the standard BMN light cone gauge of the exact S matrix approach to the quantum string $\sigma$ model \cite{Arutyunov.Frolov.Review}. In other words, the effect of these deformations at the quantum level is mysterious, in contrast to the $\beta$ deformation for example \cite{vanTongeren:2013gva}.

Recently, hints of generalised $TsT$  structures have been found also in non-abelian cases \cite{Orlando:2016qqu,Borsato.Wulff2016}. It would be interesting to try and extend our approach here, especially to the unimodular (supergravity) cases described in \cite{Borsato.Wulff2016}.

\section*{Acknowledgments}

We would especially like to thank B. Hoare for discussions, inspiring the feasibility of this approach. We would further like to thank P. Liendo and F. Loebbert for discussions, and A. A. Tseytlin for comments on the draft. S.T. is supported by L.T. The work of S.T. is supported by the Einstein Foundation Berlin in the framework of the research project "Gravitation and High Energy Physics" and acknowledges further support from the People Programme (Marie Curie Actions) of the European Union's Seventh Framework Programme FP7/2007-2013/ under REA Grant Agreement No 317089.

\newpage
\bibliographystyle{jhep}
\bibliography{References}

\providecommand{\href}[2]{#2}\begingroup\raggedright\begin{thebibliography}{10}

\bibitem{Maldacena.LargeNLimit}
J.~M. Maldacena, {\it {The Large N limit of superconformal field theories and
  supergravity}},  {\em Int. J. Theor. Phys.} {\bf 38} (1999) 1113--1133,
  [\href{http://arxiv.org/abs/hep-th/9711200}{{\tt hep-th/9711200}}]. [Adv.
  Theor. Math. Phys.2,231(1998)].

\bibitem{Arutyunov.Frolov.Review}
G.~{Arutyunov} and S.~{Frolov}, {\it {Foundations of the AdS$_{5}$ $\times$
  S$^{5}$ superstring: I}},  {\em Journal of Physics A Mathematical General}
  {\bf 42} (June, 2009) 254003, [\href{http://arxiv.org/abs/0901.4937}{{\tt
  arXiv:0901.4937}}].

\bibitem{Beisert.Review}
N.~Beisert et~al., {\it {Review of AdS/CFT Integrability: An Overview}},  {\em
  Lett. Math. Phys.} {\bf 99} (2012) 3--32,
  [\href{http://arxiv.org/abs/1012.3982}{{\tt arXiv:1012.3982}}].

\bibitem{Bombardelli.et.al.Integrability.GaugeGravity}
D.~Bombardelli, A.~Cagnazzo, R.~Frassek, F.~Levkovich-Maslyuk, F.~Loebbert,
  S.~Negro, I.~M. Szécsényi, A.~Sfondrini, S.~J. van Tongeren, and
  A.~Torrielli, {\it {An integrability primer for the gauge-gravity
  correspondence: An introduction}},  {\em J. Phys.} {\bf A49} (2016), no.~32
  320301, [\href{http://arxiv.org/abs/1606.02945}{{\tt arXiv:1606.02945}}].

\bibitem{Lunin.Maldacena.Deformation}
O.~{Lunin} and J.~{Maldacena}, {\it {Deforming field theories with U(1)
  $\times$ U(1) global symmetry and their gravity duals}},  {\em Journal of
  High Energy Physics} {\bf 5} (May, 2005) 033,
  [\href{http://arxiv.org/abs/hep-th/0502086}{{\tt hep-th/0502086}}].

\bibitem{Frolov:2005ty}
S.~Frolov, R.~Roiban, and A.~A. Tseytlin, {\it {Gauge-string duality for
  superconformal deformations of N=4 super Yang-Mills theory}},  {\em JHEP}
  {\bf 0507} (2005) 045, [\href{http://arxiv.org/abs/hep-th/0503192}{{\tt
  hep-th/0503192}}].

\bibitem{Frolov.LaxPair.Lunin-Maldacena.Background}
S.~{Frolov}, {\it {Lax pair for strings in Lunin-Maldacena background}},  {\em
  Journal of High Energy Physics} {\bf 5} (May, 2005) 069,
  [\href{http://arxiv.org/abs/hep-th/0503201}{{\tt hep-th/0503201}}].

\bibitem{Klimcik:2002zj}
C.~Klimcik, {\it {Yang-Baxter sigma models and dS/AdS T duality}},  {\em JHEP}
  {\bf 0212} (2002) 051, [\href{http://arxiv.org/abs/hep-th/0210095}{{\tt
  hep-th/0210095}}].

\bibitem{Klimcik.YBSM.Integrability}
C.~{Klim{\v c}{\'{\i}}k}, {\it {On integrability of the Yang-Baxter
  {$\sigma$}-model}},  {\em Journal of Mathematical Physics} {\bf 50} (Apr.,
  2009) 043508--043508, [\href{http://arxiv.org/abs/0802.3518}{{\tt
  arXiv:0802.3518}}].

\bibitem{DMV2}
F.~{Delduc}, M.~{Magro}, and B.~{Vicedo}, {\it {On classical q-deformations of
  integrable {$\sigma$}-models}},  {\em Journal of High Energy Physics} {\bf
  11} (Nov., 2013) 192, [\href{http://arxiv.org/abs/1308.3581}{{\tt
  arXiv:1308.3581}}].

\bibitem{DMV1}
F.~{Delduc}, M.~{Magro}, and B.~{Vicedo}, {\it {Integrable Deformation of the
  AdS$_{5}\times$S$^{5}$ Superstring Action}},  {\em Physical Review Letters}
  {\bf 112} (Feb., 2014) 051601, [\href{http://arxiv.org/abs/1309.5850}{{\tt
  arXiv:1309.5850}}].

\bibitem{Sfetsos:2013wia}
K.~Sfetsos, {\it {Integrable interpolations: From exact CFTs to non-Abelian
  T-duals}},  {\em Nucl.Phys.} {\bf B880} (2014) 225--246,
  [\href{http://arxiv.org/abs/1312.4560}{{\tt arXiv:1312.4560}}].

\bibitem{Hollowood:2014qma}
T.~J. Hollowood, J.~L. Miramontes, and D.~M. Schmidtt, {\it {An Integrable
  Deformation of the $\ads$ Superstring}},  {\em J.Phys.} {\bf A47} (2014),
  no.~49 495402, [\href{http://arxiv.org/abs/1409.1538}{{\tt
  arXiv:1409.1538}}].

\bibitem{Demulder:2015lva}
S.~Demulder, K.~Sfetsos, and D.~C. Thompson, {\it {Integrable
  $\lambda$-deformations: Squashing Coset CFTs and $AdS_5\times S^5$}},  {\em
  JHEP} {\bf 07} (2015) 019, [\href{http://arxiv.org/abs/1504.02781}{{\tt
  arXiv:1504.02781}}].

\bibitem{Klimcik:1995ux}
C.~Klimcik and P.~Severa, {\it {Dual nonAbelian duality and the Drinfeld
  double}},  {\em Phys. Lett.} {\bf B351} (1995) 455--462,
  [\href{http://arxiv.org/abs/hep-th/9502122}{{\tt hep-th/9502122}}].

\bibitem{Klimcik:1995jn}
C.~Klimcik, {\it {Poisson-Lie T duality}},  {\em Nucl. Phys. Proc. Suppl.} {\bf
  46} (1996) 116--121, [\href{http://arxiv.org/abs/hep-th/9509095}{{\tt
  hep-th/9509095}}].

\bibitem{Vicedo:2015pna}
B.~Vicedo, {\it {Deformed integrable $\sigma$-models, classical R-matrices and
  classical exchange algebra on Drinfel’d doubles}},  {\em J. Phys.} {\bf
  A48} (2015), no.~35 355203, [\href{http://arxiv.org/abs/1504.06303}{{\tt
  arXiv:1504.06303}}].

\bibitem{Hoare:2015gda}
B.~Hoare and A.~A. Tseytlin, {\it {On integrable deformations of superstring
  sigma models related to AdS$_n$ $\times$ S$^n$ supercosets}},  {\em Nucl.
  Phys.} {\bf B897} (2015) 448--478,
  [\href{http://arxiv.org/abs/1504.07213}{{\tt arXiv:1504.07213}}].

\bibitem{Sfetsos:2015nya}
K.~Sfetsos, K.~Siampos, and D.~C. Thompson, {\it {Generalised integrable
  $\lambda$- and $\eta$-deformations and their relation}},  {\em Nucl. Phys.}
  {\bf B899} (2015) 489--512, [\href{http://arxiv.org/abs/1506.05784}{{\tt
  arXiv:1506.05784}}].

\bibitem{Klimcik:2015gba}
C.~Klimcik, {\it {$\eta$ and $\lambda$ deformations as ${\cal E}$-models}},
  {\em Nucl. Phys.} {\bf B900} (2015) 259--272,
  [\href{http://arxiv.org/abs/1508.05832}{{\tt arXiv:1508.05832}}].

\bibitem{Delduc:2016ihq}
F.~Delduc, S.~Lacroix, M.~Magro, and B.~Vicedo, {\it {On q-deformed symmetries
  as Poisson-Lie symmetries and application to Yang-Baxter type models}},
  \href{http://arxiv.org/abs/1606.01712}{{\tt arXiv:1606.01712}}.

\bibitem{Borsato:2016zcf}
R.~Borsato, A.~A. Tseytlin, and L.~Wulff, {\it {Supergravity background of
  $\lambda$-deformed model for AdS$_2 \times$ S$^2$ supercoset}},  {\em Nucl.
  Phys.} {\bf B905} (2016) 264--292,
  [\href{http://arxiv.org/abs/1601.08192}{{\tt arXiv:1601.08192}}].

\bibitem{Chervonyi:2016ajp}
Y.~Chervonyi and O.~Lunin, {\it {Supergravity background of the
  $\lambda$-deformed AdS$_3 \times$ S$^3$ supercoset}},  {\em Nucl. Phys.} {\bf
  B910} (2016) 685--711, [\href{http://arxiv.org/abs/1606.00394}{{\tt
  arXiv:1606.00394}}].

\bibitem{Kawaguchi:2014qwa}
I.~Kawaguchi, T.~Matsumoto, and K.~Yoshida, {\it {Jordanian deformations of the
  $\ads$ superstring}},  {\em JHEP} {\bf 1404} (2014) 153,
  [\href{http://arxiv.org/abs/1401.4855}{{\tt arXiv:1401.4855}}].

\bibitem{Delduc:2014kha}
F.~Delduc, M.~Magro, and B.~Vicedo, {\it {Derivation of the action and
  symmetries of the $q$-deformed $\ads$ superstring}},  {\em JHEP} {\bf 1410}
  (2014) 132, [\href{http://arxiv.org/abs/1406.6286}{{\tt arXiv:1406.6286}}].

\bibitem{vanTongeren:2015uha}
S.~J. van Tongeren, {\it {Yang–Baxter deformations, AdS/CFT, and
  twist-noncommutative gauge theory}},  {\em Nucl. Phys.} {\bf B904} (2016)
  148--175, [\href{http://arxiv.org/abs/1506.01023}{{\tt arXiv:1506.01023}}].

\bibitem{Borsato.Wulff2016}
R.~Borsato and L.~Wulff, {\it {Target space supergeometry of $\eta$ and
  $\lambda$-deformed strings}},  \href{http://arxiv.org/abs/1608.03570}{{\tt
  arXiv:1608.03570}}.

\bibitem{Arutyunov:2015mqj}
G.~Arutyunov, S.~Frolov, B.~Hoare, R.~Roiban, and A.~A. Tseytlin, {\it {Scale
  invariance of the $\eta$-deformed $AdS_5\times S^5$ superstring, T-duality
  and modified type II equations}},  {\em Nucl. Phys.} {\bf B903} (2016)
  262--303, [\href{http://arxiv.org/abs/1511.05795}{{\tt arXiv:1511.05795}}].

\bibitem{Wulff:2016tju}
L.~Wulff and A.~A. Tseytlin, {\it {Kappa-symmetry of superstring sigma model
  and generalized 10d supergravity equations}},  {\em JHEP} {\bf 06} (2016)
  174, [\href{http://arxiv.org/abs/1605.04884}{{\tt arXiv:1605.04884}}].

\bibitem{Arutyunov.Borsato.Frolov.eta-deformations}
G.~{Arutyunov}, R.~{Borsato}, and S.~{Frolov}, {\it {Puzzles of
  {$\eta$}-deformed AdS$_{5}$ {$\times$} S$^{5}$}},  {\em Journal of High
  Energy Physics} {\bf 12} (Dec., 2015) 49,
  [\href{http://arxiv.org/abs/1507.04239}{{\tt arXiv:1507.04239}}].

\bibitem{Hoare:2015wia}
B.~Hoare and A.~A. Tseytlin, {\it {Type IIB supergravity solution for the
  T-dual of the $\eta$-deformed AdS$_{5} \times$ S$^{5}$ superstring}},  {\em
  JHEP} {\bf 10} (2015) 060, [\href{http://arxiv.org/abs/1508.01150}{{\tt
  arXiv:1508.01150}}].

\bibitem{Hoare.Tongeren.NonSplit.Split}
B.~Hoare and S.~J. van Tongeren, {\it {Non-split and split deformations of
  $AdS_5$}},  \href{http://arxiv.org/abs/1605.03552}{{\tt arXiv:1605.03552}}.

\bibitem{Arutynov:2014ota}
G.~Arutyunov, M.~de~Leeuw, and S.~J. van Tongeren, {\it {The exact spectrum and
  mirror duality of the $(\ads)_\eta$ superstring}},  {\em Theor.Math.Phys.}
  {\bf 182} (2015), no.~1 23--51, [\href{http://arxiv.org/abs/1403.6104}{{\tt
  arXiv:1403.6104}}].

\bibitem{Arutyunov:2014cra}
G.~Arutyunov and S.~J. van Tongeren, {\it {The $\mathrm{AdS}_5 \times
  \mathrm{S}^5$ mirror model as a string}},  {\em Phys.Rev.Lett.} {\bf 113}
  (2014) 261605, [\href{http://arxiv.org/abs/1406.2304}{{\tt
  arXiv:1406.2304}}].

\bibitem{Arutyunov:2014jfa}
G.~Arutyunov and S.~J. van Tongeren, {\it {Double Wick rotating Green-Schwarz
  strings}},  {\em JHEP} {\bf 1505} (2015) 027,
  [\href{http://arxiv.org/abs/1412.5137}{{\tt arXiv:1412.5137}}].

\bibitem{Pachol:2015mfa}
A.~Pacho{\l} and S.~J. van Tongeren, {\it {Quantum deformations of the flat
  space superstring}},  {\em Phys. Rev.} {\bf D93} (2016) 026008,
  [\href{http://arxiv.org/abs/1510.02389}{{\tt arXiv:1510.02389}}].

\bibitem{Hoare.Tongeren.Jordanian}
B.~Hoare and S.~J. van Tongeren, {\it {On jordanian deformations of $AdS_5$ and
  supergravity}},  \href{http://arxiv.org/abs/1605.03554}{{\tt
  arXiv:1605.03554}}.

\bibitem{Kyono.Yoshida.Yang-Baxter-deformation}
H.~{Kyono} and K.~{Yoshida}, {\it {Supercoset construction of Yang-Baxter
  deformed $AdS_5 \times S^5$ backgrounds}},  {\em ArXiv e-prints} (May, 2016)
  [\href{http://arxiv.org/abs/1605.02519}{{\tt arXiv:1605.02519}}].

\bibitem{Orlando:2016qqu}
D.~Orlando, S.~Reffert, J.-i. Sakamoto, and K.~Yoshida, {\it {Generalized type
  IIB supergravity equations and non-Abelian classical r-matrices}},
  \href{http://arxiv.org/abs/1607.00795}{{\tt arXiv:1607.00795}}.

\bibitem{Matsumoto:2014gwa}
T.~Matsumoto and K.~Yoshida, {\it {Integrability of classical strings dual for
  noncommutative gauge theories}},  {\em JHEP} {\bf 1406} (2014) 163,
  [\href{http://arxiv.org/abs/1404.3657}{{\tt arXiv:1404.3657}}].

\bibitem{MatsumotoYoshida.Gravity-CYBE}
T.~{Matsumoto} and K.~{Yoshida}, {\it {Integrable deformations of the AdS$_{5}
  \times $S$^{5}$ superstring and the classical Yang-Baxter equation - Towards
  the gravity/CYBE correspondence -}},  {\em Journal of Physics Conference
  Series} {\bf 563} (Nov., 2014) 012020,
  [\href{http://arxiv.org/abs/1410.0575}{{\tt arXiv:1410.0575}}].

\bibitem{Tongeren.CYBE-deformations}
S.~J.~v. {Tongeren}, {\it {On classical Yang-Baxter based deformations of the
  AdS$_{5}$ {$\times$} S$^{5}$ superstring}},  {\em Journal of High Energy
  Physics} {\bf 6} (June, 2015) 48,
  [\href{http://arxiv.org/abs/1504.05516}{{\tt arXiv:1504.05516}}].

\bibitem{Matsumoto:2016lnr}
T.~Matsumoto and K.~Yoshida, {\it {Towards the gravity/CYBE correspondence --
  $the$ $current$ $status$ --}},  {\em J. Phys. Conf. Ser.} {\bf 670} (2016),
  no.~1 012033.

\bibitem{Matsumoto:2014nra}
T.~Matsumoto and K.~Yoshida, {\it {Lunin-Maldacena backgrounds from the
  classical Yang-Baxter equation - towards the gravity/CYBE correspondence}},
  {\em JHEP} {\bf 1406} (2014) 135, [\href{http://arxiv.org/abs/1404.1838}{{\tt
  arXiv:1404.1838}}].

\bibitem{Metsaev.Tseytlin.TypeIIbGSAction.AdS5S5}
R.~R. {Metsaev} and A.~A. {Tseytlin}, {\it {Type IIB superstring action in AdS
  $_{5} \times$ S$^{5}$ background}},  {\em Nuclear Physics B} {\bf 533} (Nov.,
  1998) 109--126, [\href{http://arxiv.org/abs/hep-th/9805028}{{\tt
  hep-th/9805028}}].

\bibitem{Buscher}
T.~H. {Buscher}, {\it {A Symmetry of the String Background Field Equations}},
  {\em Physics Letters B} {\bf 194} (July, 1987) 59--62.

\bibitem{Berkovits.Maldacena.fermT-Duality}
N.~{Berkovits} and J.~{Maldacena}, {\it {Dual superconformal symmetry, and the
  amplitude/Wilson loop connection}},  {\em Journal of High Energy Physics}
  {\bf 9} (Sept., 2008) 062, [\href{http://arxiv.org/abs/0807.3196}{{\tt
  arXiv:0807.3196}}].

\bibitem{Beisert.Ricci.Tseytlin.Wolf.DualSuperconfSym}
N.~{Beisert}, R.~{Ricci}, A.~A. {Tseytlin}, and M.~{Wolf}, {\it {Dual
  superconformal symmetry from $AdS_{5}{\times}S^{5}$ superstring
  integrability}},  {\em Physical Review D} {\bf 78} (Dec., 2008) 126004,
  [\href{http://arxiv.org/abs/0807.3228}{{\tt arXiv:0807.3228}}].

\bibitem{Giveon.et.al.Target.space.Duality}
A.~{Giveon}, M.~{Porrati}, and E.~{Rabinovici}, {\it {Target space duality in
  string theory}},  {\em Physics Reports} {\bf 244} (Aug., 1994) 77--202,
  [\href{http://arxiv.org/abs/hep-th/9401139}{{\tt hep-th/9401139}}].

\bibitem{Alday:2005ww}
L.~F. Alday, G.~Arutyunov, and S.~Frolov, {\it {Green-Schwarz strings in
  TsT-transformed backgrounds}},  {\em JHEP} {\bf 06} (2006) 018,
  [\href{http://arxiv.org/abs/hep-th/0512253}{{\tt hep-th/0512253}}].

\bibitem{Siegel:1993th}
W.~Siegel, {\it {Superspace duality in low-energy superstrings}},  {\em Phys.
  Rev.} {\bf D48} (1993) 2826--2837,
  [\href{http://arxiv.org/abs/hep-th/9305073}{{\tt hep-th/9305073}}].

\bibitem{Fre.Grassi.Sommovigo.Trigiante}
P.~Fre, P.~A. Grassi, L.~Sommovigo, and M.~Trigiante, {\it {Theory of
  Superdualities and the Orthosymplectic Supergroup}},  {\em Nucl. Phys.} {\bf
  B825} (2010) 177--202, [\href{http://arxiv.org/abs/0906.2510}{{\tt
  arXiv:0906.2510}}].

\bibitem{Bergshoeff.Hull.Ortin.TDuality}
E.~{Bergshoeff}, C.~{Hull}, and T.~{Ort{\'{\i}}n}, {\it {Duality in the type-II
  superstring effective action}},  {\em Nuclear Physics B} {\bf 451} (Oct.,
  1995) 547--575, [\href{http://arxiv.org/abs/hep-th/9504081}{{\tt
  hep-th/9504081}}].

\bibitem{Kulik.Roiban.TDuality}
B.~{Kulik} and R.~{Roiban}, {\it {T-duality of the Green-Schwarz superstring}},
   {\em Journal of High Energy Physics} {\bf 9} (Sept., 2002) 007,
  [\href{http://arxiv.org/abs/hep-th/0012010}{{\tt hep-th/0012010}}].

\bibitem{Hassan.TDualityCurvedBackgrounds}
S.~F. {Hassan}, {\it {T-duality, space-time spinors and R-R fields in curved
  backgrounds}},  {\em Nuclear Physics B} {\bf 568} (Mar., 2000) 145--161,
  [\href{http://arxiv.org/abs/hep-th/9907152}{{\tt hep-th/9907152}}].

\bibitem{Fukuma.Oota.Tanaka.T-Duality.RR-Pot}
M.~{Fukuma}, T.~{Oota}, and H.~{Tanaka}, {\it {Comments on T-Dualities of
  Ramond-Ramond Potentials}},  {\em Progress of Theoretical Physics} {\bf 103}
  (Feb., 2000) 425--446, [\href{http://arxiv.org/abs/hep-th/9907132}{{\tt
  hep-th/9907132}}].

\bibitem{Brace.Morariu.Zumino.TDuality.RR-background.Matrix.Model}
D.~{Brace}, B.~{Morariu}, and B.~{Zumino}, {\it {T-duality and Ramond-Ramond
  backgrounds in the Matrix model}},  {\em Nuclear Physics B} {\bf 549} (May,
  1999) 181--193, [\href{http://arxiv.org/abs/hep-th/9811213}{{\tt
  hep-th/9811213}}].

\bibitem{Hassan.SOdd.Spinors}
S.~F. Hassan, {\it {SO(d,d) transformations of Ramond-Ramond fields and
  space-time spinors}},  {\em Nucl. Phys.} {\bf B583} (2000) 431--453,
  [\href{http://arxiv.org/abs/hep-th/9912236}{{\tt hep-th/9912236}}].

\bibitem{Sfetsos.Siampos.Thompson.FermionicTDuality}
K.~Sfetsos, K.~Siampos, and D.~C. Thompson, {\it {Canonical pure spinor
  (Fermionic) T-duality}},  {\em Class. Quant. Grav.} {\bf 28} (2011) 055010,
  [\href{http://arxiv.org/abs/1007.5142}{{\tt arXiv:1007.5142}}].

\bibitem{Hoare:2014oua}
B.~Hoare, {\it {Towards a two-parameter q-deformation of AdS$_3 \times S^3
  \times M^4$ superstrings}},  {\em Nucl. Phys.} {\bf B891} (2015) 259--295,
  [\href{http://arxiv.org/abs/1411.1266}{{\tt arXiv:1411.1266}}].

\bibitem{Patera.Winternitz.Zassenhaus.ContinuousSubgroups}
J.~Patera, P.~Winternitz, and H.~Zassenhaus, {\it Continuous subgroups of the
  fundamental groups of physics. i. general method and the poincaré group},
  {\em Journal of Mathematical Physics} {\bf 16} (1975), no.~8.

\bibitem{Hashimoto:1999ut}
A.~Hashimoto and N.~Itzhaki, {\it {Noncommutative Yang-Mills and the AdS / CFT
  correspondence}},  {\em Phys.Lett.} {\bf B465} (1999) 142--147,
  [\href{http://arxiv.org/abs/hep-th/9907166}{{\tt hep-th/9907166}}].

\bibitem{Maldacena:1999mh}
J.~M. Maldacena and J.~G. Russo, {\it {Large N limit of noncommutative gauge
  theories}},  {\em JHEP} {\bf 9909} (1999) 025,
  [\href{http://arxiv.org/abs/hep-th/9908134}{{\tt hep-th/9908134}}].

\bibitem{Fokken:2013aea}
J.~Fokken, C.~Sieg, and M.~Wilhelm, {\it {Non-conformality of ${{\gamma
  }_{i}}$-deformed N = 4 SYM theory}},  {\em J.Phys.} {\bf A47} (2014) 455401,
  [\href{http://arxiv.org/abs/1308.4420}{{\tt arXiv:1308.4420}}].

\bibitem{vanTongeren:2013gva}
S.~J. van Tongeren, {\it {Integrability of the $\ads$ superstring and its
  deformations}},  {\em J.Phys.} {\bf A47} (2014), no.~43 433001,
  [\href{http://arxiv.org/abs/1310.4854}{{\tt arXiv:1310.4854}}].

\bibitem{Fokken:2014soa}
J.~Fokken, C.~Sieg, and M.~Wilhelm, {\it {A piece of cake: the ground-state
  energies in $\gamma_{i}$ -deformed $ \mathcal{N} $ = 4 SYM theory at leading
  wrapping order}},  {\em JHEP} {\bf 1409} (2014) 78,
  [\href{http://arxiv.org/abs/1405.6712}{{\tt arXiv:1405.6712}}].

\bibitem{Arutyunov:2014cda}
G.~Arutyunov and D.~Medina-Rincon, {\it {Deformed Neumann model from spinning
  strings on ($AdS_5 \times S^5$)$_\eta$}},  {\em JHEP} {\bf 10} (2014) 050,
  [\href{http://arxiv.org/abs/1406.2536}{{\tt arXiv:1406.2536}}].

\bibitem{Banerjee:2014bca}
A.~Banerjee and K.~L. Panigrahi, {\it {On the rotating and oscillating strings
  in (AdS$_{3}$ x S$^{3}$)$_{\kappa}$}},  {\em JHEP} {\bf 09} (2014) 048,
  [\href{http://arxiv.org/abs/1406.3642}{{\tt arXiv:1406.3642}}].

\bibitem{Kameyama:2014vma}
T.~Kameyama and K.~Yoshida, {\it {A new coordinate system for $q$-deformed
  AdS$_{5} \times$ S$^5$ and classical string solutions}},  {\em J. Phys.} {\bf
  A48} (2015), no.~7 075401, [\href{http://arxiv.org/abs/1408.2189}{{\tt
  arXiv:1408.2189}}].

\bibitem{Arutyunov:2016ysi}
G.~Arutyunov, M.~Heinze, and D.~Medina-Rincon, {\it {Integrability of the
  eta-deformed Neumann-Rosochatius model}},
  \href{http://arxiv.org/abs/1607.05190}{{\tt arXiv:1607.05190}}.

\bibitem{Banerjee:2016xbb}
A.~Banerjee and K.~L. Panigrahi, {\it {On circular strings in $(AdS_3 \times
  S^3)_{\varkappa}$}},  \href{http://arxiv.org/abs/1607.04208}{{\tt
  arXiv:1607.04208}}.

\end{thebibliography}\endgroup

\end{document}